\documentclass[10pt,aps,prd,twocolumn,superscriptaddress,nofootinbib,floatfix]{revtex4-2}
\usepackage{graphicx}
\usepackage{bm}
\usepackage{dcolumn}
\usepackage{amsmath,amsthm,amsfonts,amssymb}
\usepackage{mathtools}
\usepackage{appendix}
\usepackage{xcolor}
\usepackage{orcidlink}
\usepackage{hyperref}
\hypersetup{
    colorlinks=true,
    linkcolor=red,
    citecolor=red,
    urlcolor=magenta
}

\begin{document}

\title{Topologically Charged Morris-Thorne-type Wormholes and the Energy Conditions}

\author{ Faizuddin Ahmed\orcidlink{0000-0003-2196-9622}}
\email[E-mail: ]{fahmed@rgu.ac ; faizuddin@associates.iucaa.in}
\affiliation{Department of Physics, The Assam Royal Global University, Guwahati, 781035, Assam, India}

\author{Md Sabir Ali\orcidlink{0000-0001-6670-7955}}
\email[E-mail: ]{alimd.sabir3@gamil.com; sabirali@mahishadalrajcollege.ac.in}
\affiliation{Department of Physics, Mahishadal Raj College, West Bengal, 721628, India}

\author{Adnan Malik\orcidlink{0000-0002-6034-0117}}
\email[E-mail: ]{adnan.malik@zjnu.edu.cn}
\affiliation{Center for Theoretical Physics, Khazar University, 41 Mehseti Str., AZ1096 Baku, Azerbaijan}
\affiliation{Jadara Research Center Jadara University, Irbid 21110, Jordan}

\begin{abstract}
{\small In this paper, we investigate topologically charged Morris--Thorne-type traversable wormholes by solving the Einstein field equations in the presence of an anisotropic fluid source and analysing the resulting wormhole solutions. We consider several classes of shape functions, namely, $A(r)=r_0 e^{r_0-r}$, $A(r)=r_0\,a^{r-r_0}$ with $0<a<1$, $A(r)=r_0\left(\cosh r_0/\cosh r\right)^{\delta}$ with $\delta\geq1$, $A(r)=1/r+\ln(r/r_0)$, $A(r)=Br^n+(1-B)$, $A(r)=r_0\,\ln(1+r)/\ln(1+r_0)$, and $A(r)=r_0+a r_0\left[(r/r_0)^{\beta}-1\right]$ with $\beta<1$ and $0<a\beta<1$, which extend the choices considered in [Eur.\ Phys.\ J.\ C \textbf{84} (2024) 1037]. For these configurations, we examine the null, weak, strong, and dominant energy conditions and analyse the influence of the topological charge on their behaviour. Furthermore, we compute the anisotropy parameter to determine whether the corresponding wormhole geometries exhibit attractive or repulsive gravitational effects. Our analysis reveals that the energy density of the anisotropic fluid remains positive for all considered configurations. However, while some energy conditions are satisfied in certain regions of the spacetime, others are violated, indicating the non-trivial role of the topological charge in controlling the physical properties and viability of these traversable wormhole solutions.\\

{\bf keywords:} traversable wormholes; global monopoles; energy conditions

}
\end{abstract}

\maketitle

\small

\section{Introduction}

Traversable wormholes are among the most intriguing and mathematically elegant solutions to Einstein's field equations. The hypothetical spacetime structures establish nontrivial topological bridges connecting either causally disconnected regions of the same universe or entirely distinct universes. Therefore, they provide the potential shortcuts through the fabric of spacetime. The modern theoretical prediction for traversable wormholes was established by Morris and Thorne \cite{Morris1988,MorrisThorne1988}, who demonstrated that suitably engineered geometries could, in principle, allow the safe passage of observers without encountering any singular structures, e.g., curvature singularities or event horizons. A follow-up paper, as investigated by Morris, Thorne, and Yurtsever, demonstrated that the traversable wormholes indeed allow the construction of closed timelike trajectories, thereby correlating the physics of the wormholes with causality conditions and time machines \cite{Morris1988}. After their proposal, traversable wormholes have become one of the central objects of investigation in gravitational physics, not only because they challenge our basic understanding of spacetime geometry, but also because of their profound implications for relativistic astrophysics, cosmology, and quantum gravity. In contrast to black holes, whose event horizon is constrained by a unidirectional surface called the event horizon, the traversable wormholes remain horizon-free, thereby permitting bidirectional communication and the exchange of matter and information between the interconnected regions without any interruption. These important characteristics have made traversable wormholes one of the most important theoretical frameworks, thereby exploring the interplay among gravitation, topology, and fundamental aspects of the spacetime fabric.

The Morris-Thorne wormholes are represented by two characteristic functions, namely, the redshift function and the shape function. The redshift function determines the gravitational redshift and has a finite value at all spacetime points to avoid event horizons. On the other hand, the shape function is related to the spatial geometry and wormhole throat. In addition, the throat must satisfy the flare-out conditions to ensure the wormhole remains open and traversable. However, within the framework of classical general relativity, this condition requires violating the null energy condition (NEC) \cite{Hochberg1997}. Under such circumstances, the matter comprising the wormhole must possess a negative energy density or sufficiently high negative radial pressure, commonly referred to as exotic matter. Since the conventional matter fields do not support such requirements, the construction of realistic traversable wormholes remains one of the principal challenges. Numerous authors have been studied wormhole solutions in general relativity (see, for examples, Refs.~\cite{Yousaf2025,Molla2025,Mustafa2025,Khatri2025,Sarkar2024,Ashraf2026,Ahmed2025,Konoplya2022,Bronnikov2022,Bronnikov2019,Maldacena2023,Churilova2021} and related references therein). However, the unconventional matter sources supporting wormhole geometries play a pivotal role in the context of modified theories of gravity. Significant progress has been achieved in modified theories of gravity, including the $f(R)$ gravity \cite{Lobo2009, Harko2011}, $f(Q)$-gravity \cite{Tayde2024,Pradhan2024,Chowdhury2025,Banerjee2021}, $f(R, T)$-gravity \cite{Rahaman2024,Mondal2024}, Einstein–Gauss–Bonnet gravity \cite{Mehdizadeh:2015jra, Kokubu:2020lxs}, Lovelock gravity \cite{Giribet:2019dmg}, scalar-tensor theories \cite{Bakopoulos:2021liw}, Horndeski gravity \cite{Mironov:2018uou}, Teleparallel gravity \cite{Tefo:2018rej}, Einstein-Cartan theory \cite{Bronnikov:2015pha,Mehdizadeh:2017dhb,Sarkar:2024efj}, and braneworld models \cite{Wong:2011pt,Antonini:2024bbm}. In many of these theories, the higher-order curvature terms contribute to the effective energy-momentum tensor, permitting ordinary matter to inflate wormholes satisfying the energy conditions while the geometric part leads to necessary violations. For detailed information regarding wormhole geometries in the modified gravity theories, see, for example, Refs.~\cite{Lobo2017, Visser1995} and references therein.

Of particular interest, in this work, we incorporate topological charge, e.g., the global monopole, which naturally occurs in spontaneous symmetry breaking \cite{Vilenkin2000}. It is argued that a global monopole charge pertaining to a spontaneous breakdown of the global $O(3)$ symmetry to $U(1)$ is produced during the early inflationary epoch of the phase transitions of the universe \cite{Barriola1989}. When we add the global monopole charge as a possible source of the energy-momentum tensor in Einstein's field equations, we have the global monopole charge-inspired traversable wormholes. These global monopoles often introduce additional geometric structures that sufficiently modify the spacetime curvature and dynamical behavior of the gravitational field. In addition, these charges can alter thermodynamic behavior, modify gravitational lensing, and influence the stability of compact objects. Such developments inspire us to investigate further whether the topological charges similarly support traversable wormholes while minimizing the effect of exotic matter. The interplay between spacetime topology and gravitation has become increasingly important in modern theoretical frameworks. The topological invariants play a crucial role in understanding several aspects of quantum gravity, including Euclidean path-integral formulations, topological quantum field theories, holography, and Ad/CFT correspondence. On the same token, the topological defects produced during the early inflationary epoch of cosmological phase transitions can significantly affect the geometry and evolution of the universe. The gravitational dynamics influenced by these topological defects naturally lead to topologically charged traversable wormholes, which can be regarded as a natural extension of the existing Morris–Thorne wormholes.

Inevitably, energy conditions, namely, Null Energy Condition (NEC), Weak Energy Condition (WEC), Dominant Energy Condition (DEC), and Strong Energy Condition (SEC), imposed on the matter field are equally important while studying the wormhole physics. These energy conditions play central roles in singularity theorems, gravitational collapse, and black hole thermodynamics \cite{Hawking1973,Wald1984}. Classically, a traversable wormhole violates the NEC in the vicinity of its throat. However, numerous studies have shown that modified theories of gravity, anisotropic matter fields, quantum corrections, and nonminimal matter--gravity couplings can significantly alleviate these violations \cite{Lobo2005,Kar2006}. Hence, understanding how topological charges influence the effective energy-momentum tensor is essential for analyzing the physical constraints for wormhole geometries. 

The recent era is celebrated as the dawn of precision gravitational astronomy, which enhances the study of wormhole geometries. The direct detection of gravitational waves from black hole mergers by LIGO/Virgo and KAGRA \cite{Abbott2016Observation,Abbott2016Properties} and horizon-scale imaging of supermassive black holes M$87$ and SgrA$^*$ by the Event Horizon Telescope (EHT) \cite{EHT2019_I,EHT2019_IV,EHT2019_VI,EHT2022_I,EHT2022_IV,EHT2022_VI} opened a window into an era of precision strong gravity, enabling exceptional tests of gravitational theories in the extreme-curvature regime. Topologically charged traversable wormholes may reconcile with the observational signatures distinct from the classical black hole geometries, including black hole silhouettes and strong lensing properties, gravitational echoes, and quasinormal spectra. These observational differences may potentially distinguish wormholes as an active area of contemporary research and motivate exploring new exact wormhole solutions beyond standard general relativity.  

Motivated by these recent developments, in this work we investigate topologically charged Morris--Thorne-type wormhole spacetimes by considering several well-established shape functions proposed in the literature, different from those examined in Ref.~\cite{EPJC2024}. For each choice of the shape function, we derive the corresponding Einstein field equations and analyze the resulting wormhole solutions assuming an anisotropic fluid as the matter source. Our analysis demonstrates that the inclusion of the global monopole parameter plays a crucial role in controlling the energy conditions. In particular, for suitable choices of the model parameters, the obtained wormhole configurations satisfy the null, weak, dominant, and strong energy conditions, thereby eliminating the need for exotic matter. We further investigate the gravitational nature of the solutions and show that, depending on the adopted shape function, the resulting wormholes can exhibit either attractive or repulsive behavior.


\section{Topologically charged wormholes and the Energy Conditions}\label{sec:2}

In this section, we present a metric anstaz representing a topologically charged Morris-Thorne-type wormhole and obtain the field equations. Therefore, we begin this part by introducing a line-element describing this wormhole space-time in ``Schwarzschild coordinate" $(t, r, \theta, \phi)$ is given by \cite{MorrisThorne1988,Morris1988,EPJC2024}
\begin{equation}
    ds^2=-e^{2\,\Phi(r)}\,dt^2+\frac{dr^2}{\alpha^2\,\left(1-A(r)/r\right)}+r^2\,(d\theta^2+\sin^2\theta\,d\phi^2),\label{b1}   
\end{equation}
where $\alpha=(1-8\,\pi\,G\,\eta^2)$ is the global monopole parameter with $\eta$ is the energy scale of the symmetry breaking. Here, the parameter $\alpha$ is introduced analogue to the topologically charged wormhole models \cite{Ahmed2023,Ahmed2023JCAP,Ahmed2023JCAP082,Ahmed2024APB} and global monopole space-time \cite{Barriola1989,Bezerra2001,Ahmed2022}. The physics associated with the global monopole space-time has been discussed in details in \cite{Bezerra2001}.

As mentioned earlier, to prevent the formation of event horizons, the red-shift function $\Phi(r)$ should be finite throughout everywhere. A particular case which we are interested here is the solution with a constant red-shift function, $\Phi'(r)=0$. Without a loss of generality, we have considered $\Phi(r)=0$. Hence the line-element (\ref{b1}) becomes
\begin{equation}
    ds^2=-dt^2+\frac{dr^2}{\alpha^2\,\left(1-A(r)/r\right)}+r^2\,(d\theta^2+\sin^2\theta\,d\phi^2).\label{b2}   
\end{equation}

The non-zero components of the Einstein’s tensor $G_{\mu\nu}$ for the metric (\ref{b2}) are given by
\begin{eqnarray}
    &&G^{t}_{t}=-\frac{(1-\alpha^2+\alpha^2\,A')}{r^2},\nonumber\\
    &&G^{r}_{r}=\frac{\alpha^2\,\left(1-\frac{A}{r}\right)-1}{r^2},\nonumber\\
    &&G^{\theta}_{\theta}=\frac{\alpha^2\,(A-r\,A')}{2\,r^3}=G^{\phi}_{\phi},\label{b4}
\end{eqnarray}
where $(')$ denotes ordinary derivative w. r. t. argument $r$. The Ricci scalar $R=g^{\mu\nu}\,R_{\mu\nu}$ is given by
\begin{equation}
    R=\frac{2\,(1-\alpha^2+\alpha^2\,A')}{r^2}.\label{b5} 
    \end{equation}

Since the space-time (\ref{b2}) is a non-vacuum solution of the field equations with the non-zero components of the Einstein tensor given in Eq. (\ref{b4}), we consider anisotropic fluid as matter content whose energy-momentum tensor is given by 
\begin{equation}
    T^{\mu\nu}=(\rho+p_t)\,U^{\mu}\,U^{\nu}+(p_r-p_t)\, \eta^{\mu}\,\eta^{\nu}+p_t\,g^{\mu\nu},\label{b6}
\end{equation}
where $\rho$ is the energy density, $p_r$ is the radial pressure and $p_t$ is the tangential pressure, respectively. Here, $U^{\mu}$ is the velocity four-vector which takes the values $U^{\mu}=(1,0,0,0)$ such that $U^{\mu}\,U_{\mu}=-1$ and $\eta^{\mu}$ is the radial vector which takes the form $\eta^{\mu}=\left(0,\alpha\,\left(1-\frac{A}{r}\right)^{1/2},0,0\right)$ such that $\eta^{\mu}\,\eta_{\mu}=1$. Also, these vector field satisfies the relation $U^{\mu}\,\eta_{\mu}=0$.

Solving the Einstein Field Equation $G_{\mu\nu}=T_{\mu\nu}$ using equations (\ref{b4}) and (\ref{b6}), we obtain the following physical quantities
\begin{eqnarray}
    &&\rho=\frac{(1-\alpha^2+\alpha^2\,A')}{r^2},\quad
    p_r=\frac{\alpha^2\,\left(1-\frac{A}{r}\right)-1}{r^2},\nonumber\\
    &&p_t=\frac{\alpha^2\,(A/r-A')}{2\,r^2},\quad \rho+p_r+2\,p_t=0.\label{b7}
\end{eqnarray}

Equation (\ref{b7}) provides the expression for the energy density, the radial pressure and the tangential pressure respectively of the fluid. We see that in the limit $\alpha \to 1$, one will find the results similar to those obtained in Morris-Thorne wormhole model \cite{MorrisThorne1988,Morris1988,Visser1995}. However, in this analysis, we are mainly interested on $\alpha \neq 1$ and show how this parameter $\alpha$ control the energy conditions as well as other quantities, such as the anisotropy parameter $\Delta$ for different types of the shape functions known in the literature.

The geometrical properties of wormholes are dependent on the shape function. In literature, various shape functions $A(r)$ are defined and wormhole structures are analyzed. Below, we consider a few known shape functions $A(r)$ and examine wormhole models and to check for the validation of the energy conditions for each wormhole model in the presence of global monopole charge.

\begin{figure*}
    \centering
    \includegraphics[width=0.3\linewidth]{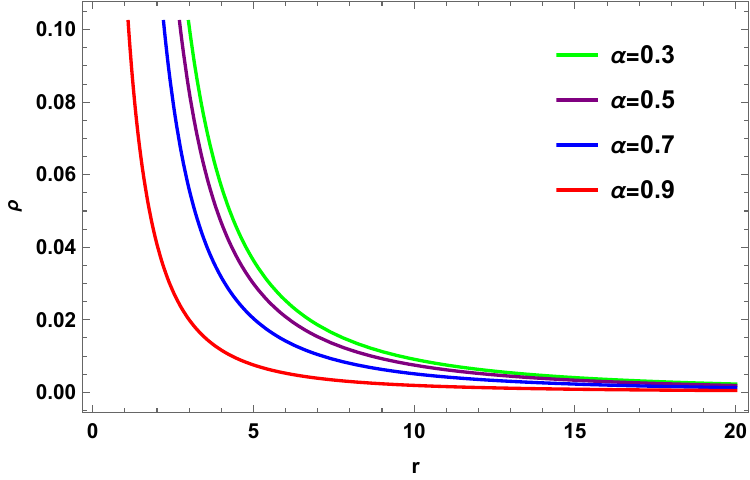}
    \includegraphics[width=0.3\linewidth]{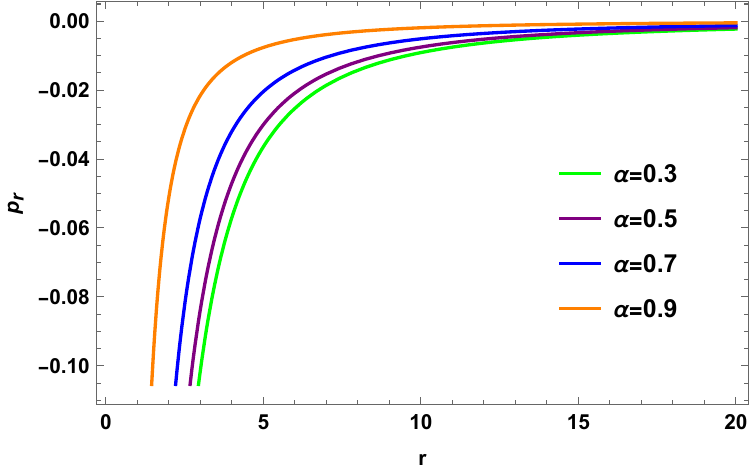}
    \includegraphics[width=0.28\linewidth]{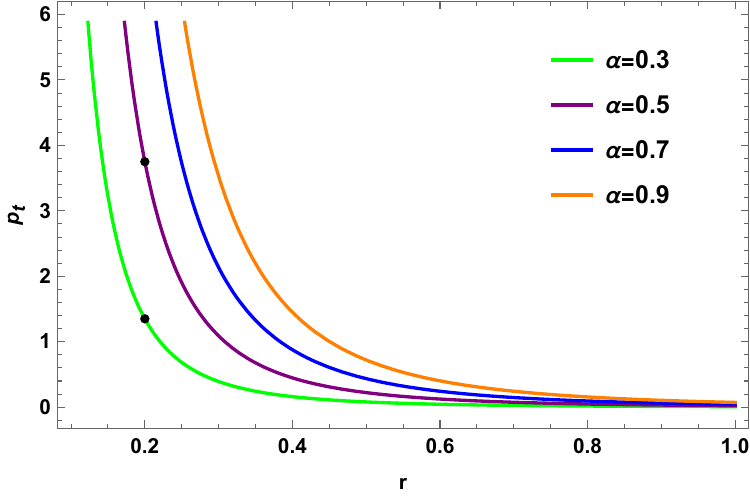}
    \caption{Behaviour of the physical quantities $\rho, p_r, p_t$ as a function of $r$ for different values of $\alpha$ for the shape function $A(r)=r_0\,\exp(r_0-r)$ with throat radius $r_0=0.2$.}
    \label{fig:1}
\end{figure*}

\subsection{Shape Function:\, $A(r)=r_0\,e^{-(r-r_0)}$.}\label{subsec:1}

In this model, we consider the following form of the shape function, as discussed in \cite{Samanta2020}, to examine the topologically charged wormhole space-time given by:
\begin{equation}
    A(r)=r_0\,e^{-(r-r_0)},\label{b51}
\end{equation}
where $r_0$ is the throat radius. We see at the throat radius $r=r_0$, the shape function becomes $A(r_0)=r_0$.

Therefore, using this shape function (\ref{b51}), we finds the energy-density, the radial pressure, and the tangential pressure from Eq. (\ref{b7}) as follows:
\begin{eqnarray}
    &&\rho=\frac{1}{r^2}\Big[1-\alpha^2-\alpha^2\,r_0\,e^{-(r-r_0)}\Big],\nonumber\\
    &&p_r=\frac{\alpha^2}{r^2}\Big[1-\frac{r_0}{r}\,e^{-(r-r_0)}\Big]-\frac{1}{r^2},\nonumber\\
    &&p_t=\frac{r_0\,\alpha^2}{2\,r^3}\,(1+r)\,e^{-(r-r_0)}.\label{b52}
\end{eqnarray}

\begin{figure*}[ht!]
    \includegraphics[width=0.3\linewidth]{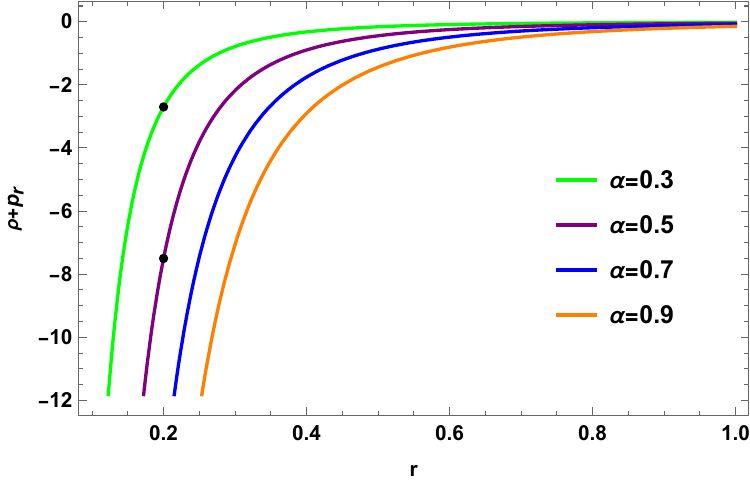}\quad\quad
    \includegraphics[width=0.3\linewidth]{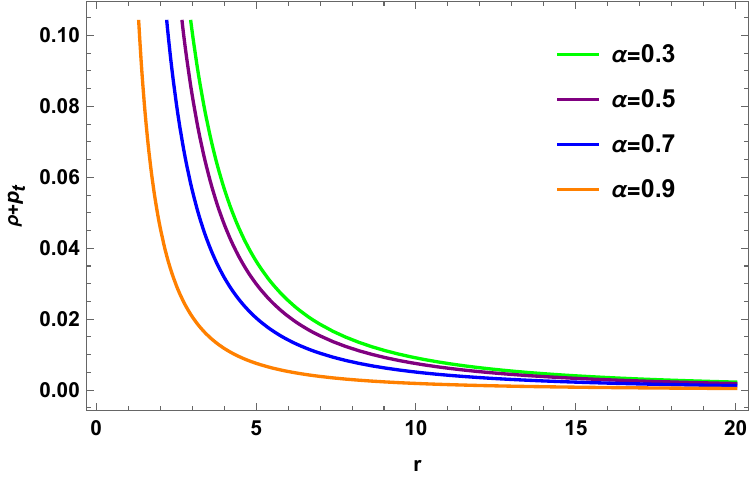}
    \caption{The weak energy condition $\rho+p_r$ (left one) and $\rho+p_t$ (right one) for the shape function $A(r)=r_0\,\exp(r_0-r)$ with throat radius $r_0=0.2$.}
    \label{fig:2}
    \hfill\\
    \includegraphics[width=0.3\linewidth]{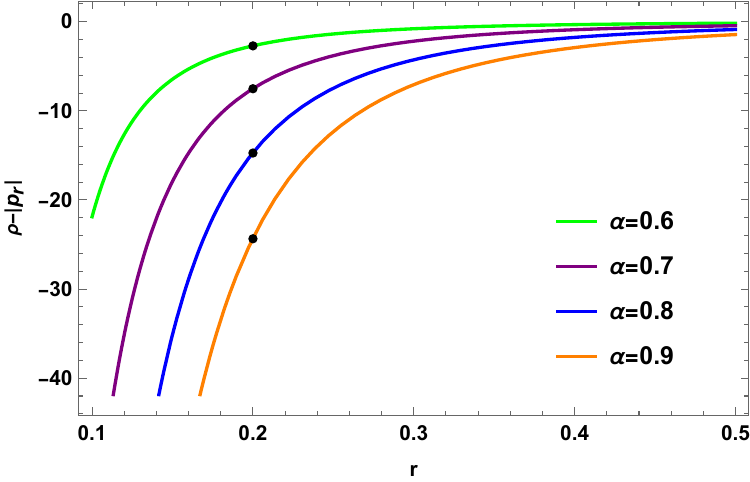}\quad\quad
    \includegraphics[width=0.3\linewidth]{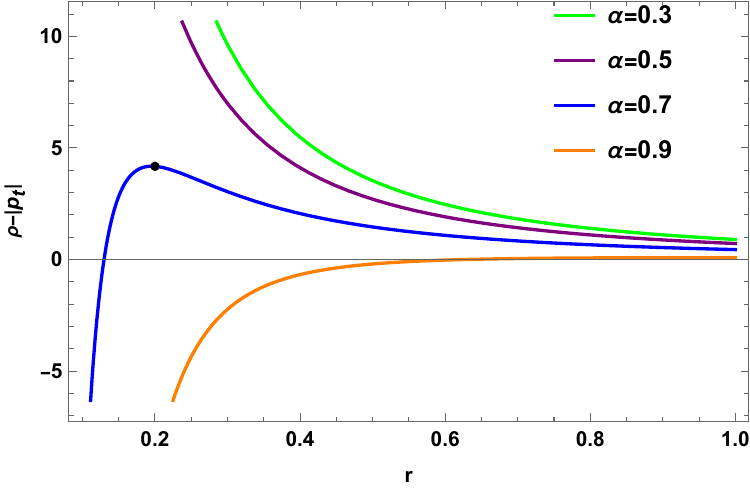}
    \caption{The behaviour of $\rho-|p_r|$ and $\rho-|p_t|$ for the shape function $A(r)=r_0\,\exp(r_0-r)$ with throat radius $r_0=0.2$.}
    \label{fig:3}
\end{figure*}

\begin{figure}[ht!]
    \centering
    \includegraphics[width=0.6\linewidth]{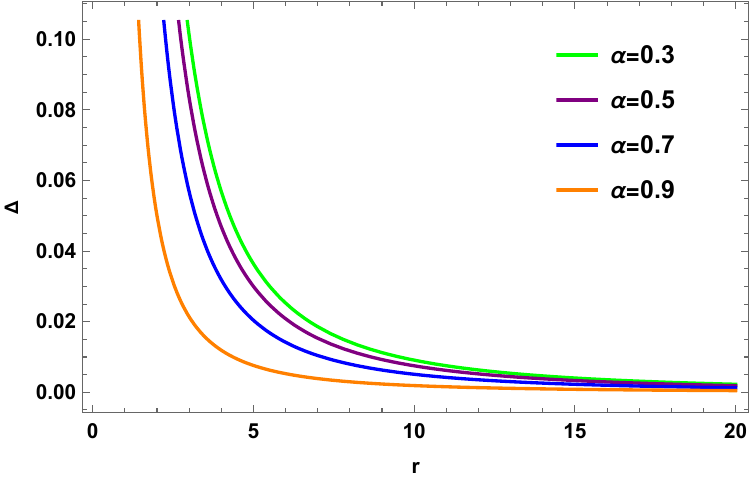}
    \caption{The nature of the anisotropy parameter $\Delta$ as a function of $r$ for the shape function $A(r)=r_0\,\exp(r_0-r)$ with throat radius $r_0=0.2$.}
    \label{fig:4}
\end{figure}

Now, we analyze the energy conditions associated with the matter-energy tensor. The WECs states that
\begin{eqnarray}
    \text{WEC}_r:\quad &&\rho+p_r=-\frac{\alpha^2\,r_0}{r^2}\,\left(1+\frac{1}{r}\right)\,e^{-(r-r_0)}<0,\nonumber\\
    \text{WEC}_t:\quad &&\rho+p_t=\frac{1-\alpha^2}{r^2}+\frac{\alpha^2\,r_0}{2\,r^3}\,(1-r)\,e^{-(r-r_0)}.\label{b53}
\end{eqnarray}

Furthermore, DECs state that
\begin{eqnarray}
    \text{DEC}_r:&&\rho-|p_r|=\frac{1}{r^2}\Big[1-\alpha^2-\alpha^2\,r_0\,e^{-(r-r_0)}\Big]\nonumber\\
    &&-\left|\frac{\alpha^2}{r^2}\Big[1-\frac{r_0}{r}\,e^{-(r-r_0)}\Big]-\frac{1}{r^2}\right|,\nonumber\\
    \text{DEC}_t:&&\rho-|p_t|=\frac{1-\alpha^2}{r^2}-\frac{\alpha^2\,r_0}{2\,r^3}\,(1+2\,r)\,e^{-(r-r_0)}.\quad\label{b54}
\end{eqnarray}

Finally, we determine the anisotropy parameter given by
\begin{equation}
    \Delta=\frac{1-\alpha^2}{r^2}+\frac{3}{2}\,\frac{\alpha^2\,r_0}{r^3}\,e^{-(r-r_0)}>0.\label{b56}
\end{equation}

\subsection{Shape Function::\, $A(r)=r_0\,\frac{a^{r}}{a^{r_0}}$,\quad $0 < a < 1$.}\label{subsec:2}

Following \cite{Mishra2021}, we adopt the following form of the shape function to study the geometry of a topologically charged wormhole spacetime:
\begin{equation}
    A(r)=r_0\,\frac{a^r}{a^{r_0}}=r_0\,\frac{e^{r\,\mbox{ln}\, a}}{e^{r_0\,\mbox{ln}\, a}}=r_0\,e^{\mathrm{b}\,(r-r_0)},\quad\quad \mathrm{b}=\mbox{ln}\,a<0,\label{b57}
\end{equation}
where $r_0$ is the throat radius. Therefore, using this shape function (\ref{b51}), we find the energy-density, the radial pressure, and the tangential pressure from Eq. (\ref{b7}) as follows:
\begin{eqnarray}
    &&\rho=\frac{1}{r^2}\Big[1-\alpha^2+\alpha^2\,\mathrm{b}\,r_0\,e^{\mathrm{b}\,(r-r_0)}\Big]>0,\nonumber\\
    &&p_r=\frac{\alpha^2}{r^2}\Big[1-\mathrm{b}\,\frac{r_0}{r}\,e^{\mathrm{b}\,(r-r_0)}\Big]-\frac{1}{r^2},\nonumber\\
    &&p_t=\frac{r_0\,\alpha^2}{2\,r^3}\,(1-\mathrm{b}\,r)\,e^{\mathrm{b}\,(r-r_0)}.\label{b58}
\end{eqnarray}

\begin{figure*}[ht!]
    \centering
    \includegraphics[width=0.3\linewidth]{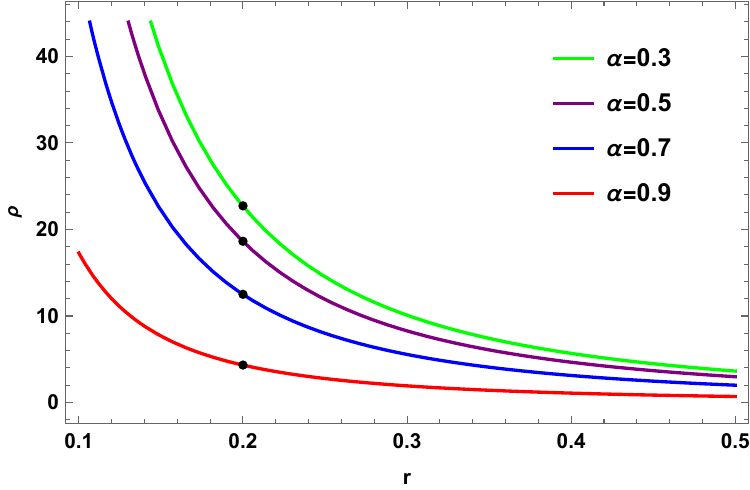}\quad
    \includegraphics[width=0.3\linewidth]{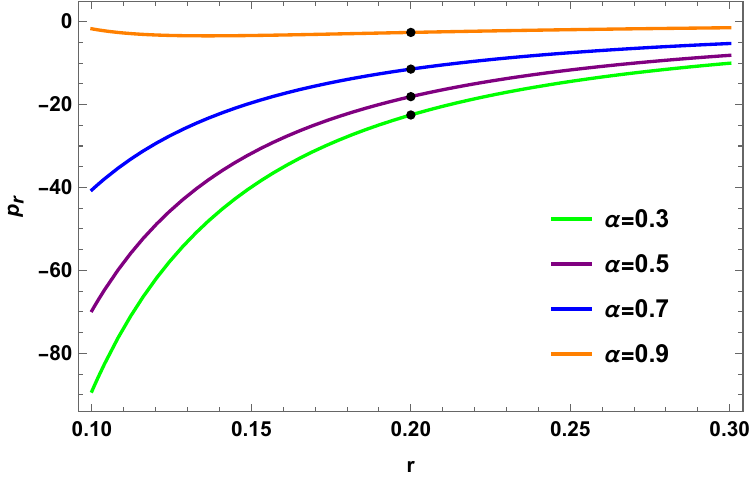}\quad
    \includegraphics[width=0.3\linewidth]{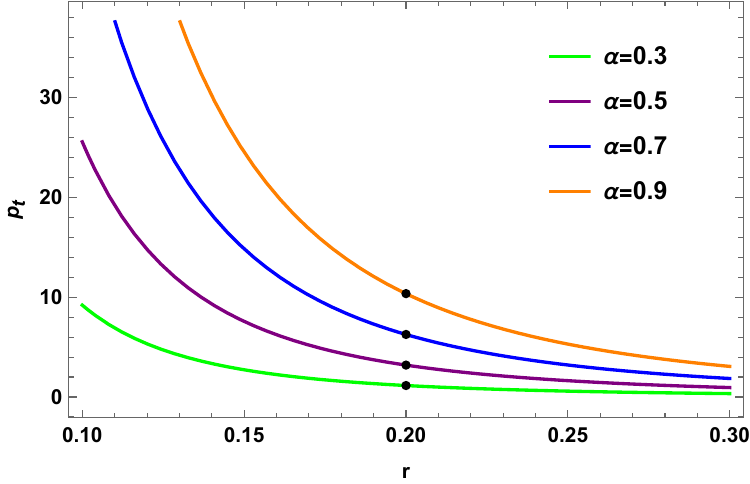}
    \caption{The behavior of physical quantities $\rho, p_r, p_t$ as a function of $r$ for different values of $\alpha$ for the shape function $A(r)=r_0\,\frac{a^{r}}{a^{r_0}}$ with throat radius $r_0=0.2$ and $a=0.9<1$.}
    \label{fig:5}
\end{figure*}

\begin{figure*}[ht!]
    \includegraphics[width=0.3\linewidth]{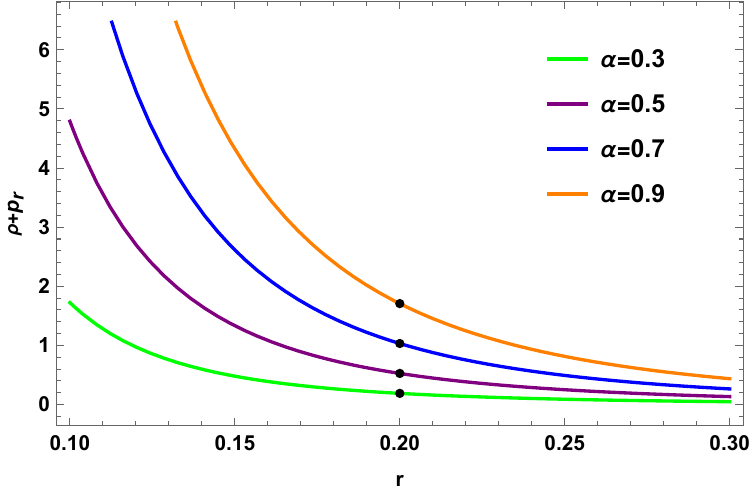}\quad\quad
    \includegraphics[width=0.3\linewidth]{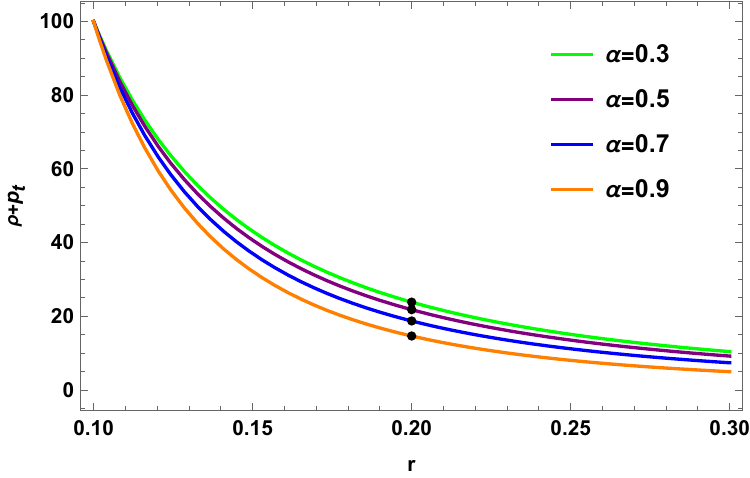}
    \caption{The weak energy condition $\rho+p_r$ (left one) and $\rho+p_t$ (right one) for the shape function $A(r)=r_0\,\frac{a^{r}}{a^{r_0}}$ with throat radius $r_0=0.2$ and $a=0.9<1$.}
    \label{fig:6}
    \hfill\\
    \includegraphics[width=0.3\linewidth]{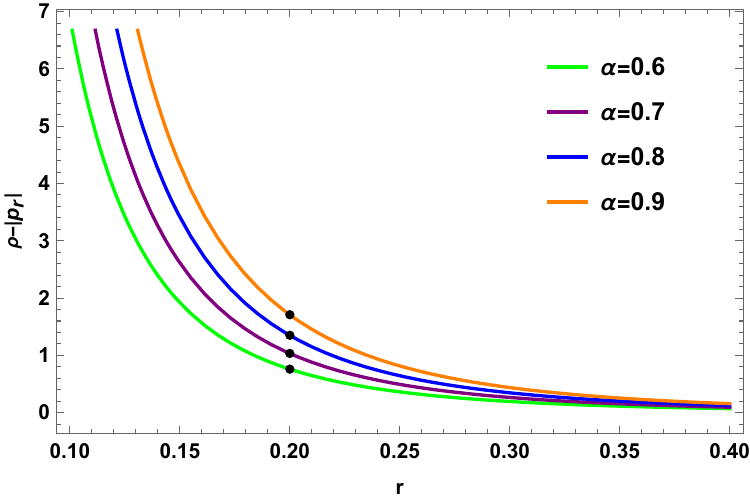}\quad\quad
    \includegraphics[width=0.3\linewidth]{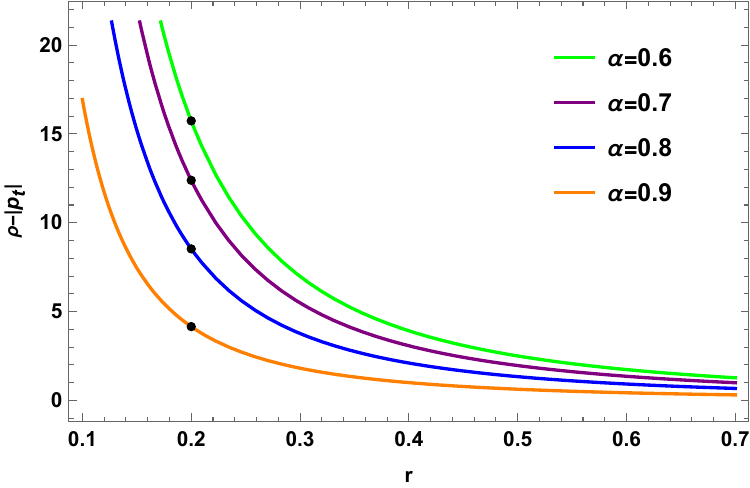}
    \caption{The behaviour of $\rho-|p_r|$ and $\rho-|p_t|$ for the shape function $A(r)=r_0\,\frac{a^{r}}{a^{r_0}}$ with throat radius $r_0=0.2$ and $a=0.9<1$.}
    \label{fig:7}
\end{figure*}

\begin{figure}[ht!]
    \begin{centering}
    \includegraphics[width=0.6\linewidth]{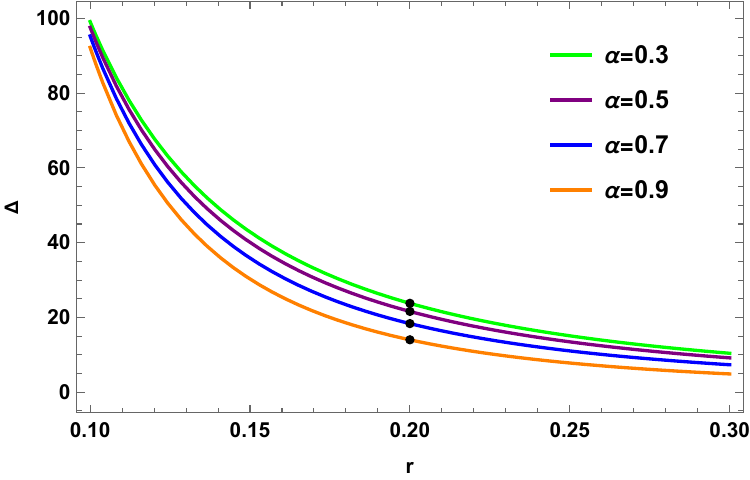}
    \par\end{centering}
    \caption{The nature of anisotropy parameter $\Delta$ as a function of $r$ for the shape function $A(r)=r_0\,\frac{a^{r}}{a^{r_0}}$ with throat radius $r_0=0.2$ and $a=0.9<1$.}
    \label{fig:8}
\end{figure}

The WECs states that
\begin{eqnarray}
    \text{WEC}_r:&&\rho+p_r=\frac{\alpha^2\,\mathrm{b}\,r_0}{r^3}\,(r-1)\,e^{\mathrm{b}\,(r-r_0)},\nonumber\\
    \text{WEC}_t:&&\rho+p_t=\frac{1-\alpha^2}{r^2}+\frac{\alpha^2\,r_0}{2\,r^3}\,(1+\mathrm{b}\,r)\,e^{\mathrm{b}\,(r-r_0)}>0.\label{b59}
\end{eqnarray}

The DECs become
\begin{eqnarray}
    \text{DEC}_r:&&\rho-|p_r|=\frac{1}{r^2}\Big[1-\alpha^2+\alpha^2\,\mathrm{b}\,r_0\,e^{\mathrm{b}\,(r-r_0)}\Big]\nonumber\\
    &&-\left|\frac{\alpha^2}{r^2}\Big[1-\mathrm{b}\,\frac{r_0}{r}\,e^{\mathrm{b}\,(r-r_0)}\Big]-\frac{1}{r^2}\right|,\nonumber\\
    \text{DEC}_t:&&\rho-|p_t|=\frac{1}{r^2}\Big[1-\alpha^2+\alpha^2\,\mathrm{b}\,r_0\,e^{\mathrm{b}\,(r-r_0)}\Big]\nonumber\\
    &&-\left|\frac{r_0\,\alpha^2}{2\,r^3}\,(1-\mathrm{b}\,r)\,e^{\mathrm{b}\,(r-r_0)}\right|.\label{b60}
\end{eqnarray}

\subsection{Shape Function:\,$A(r)=r_0\,\left(\frac{\cosh r_0}{\cosh r}\right)^{\delta}$.}\label{subsec:3}

In this model, we consider the following form of the shape function \cite{Chaudhary2024} to examine the topologically charged wormhole space-time given by
\begin{equation}
    A(r)=r_0\,\left(\frac{\cosh r_0}{\cosh r}\right)^{\delta},\quad \delta \geq 1.\label{mm1}
\end{equation}
Using this shape function (\ref{mm1}), we find the energy-density, the radial pressure, and the tangential pressure from Eq. (\ref{b7}) as follows:
\begin{eqnarray}
    &&\rho=\frac{1-\alpha^2}{r^2}-\frac{\alpha^2\,\delta\,r_0}{r^2}\,\left(\frac{\cosh r_0}{\cosh r}\right)^{\delta}\,\tanh r,\nonumber\\
    &&p_r=\frac{\alpha^2-1}{r^2}-\frac{\alpha^2}{r^2}\,\frac{r_0}{r}\,\left(\frac{\cosh r_0}{\cosh r}\right)^{\delta},\nonumber\\
    &&p_t=\frac{\alpha^2}{2\,r^2}\,\Big[\frac{r_0}{r}+\delta\,r_0\,\tanh r\Big]\,\left(\frac{\cosh r_0}{\cosh r}\right)^{\delta},\label{mm2}
\end{eqnarray}

\begin{figure*}
    \centering
    \includegraphics[width=0.3\linewidth]{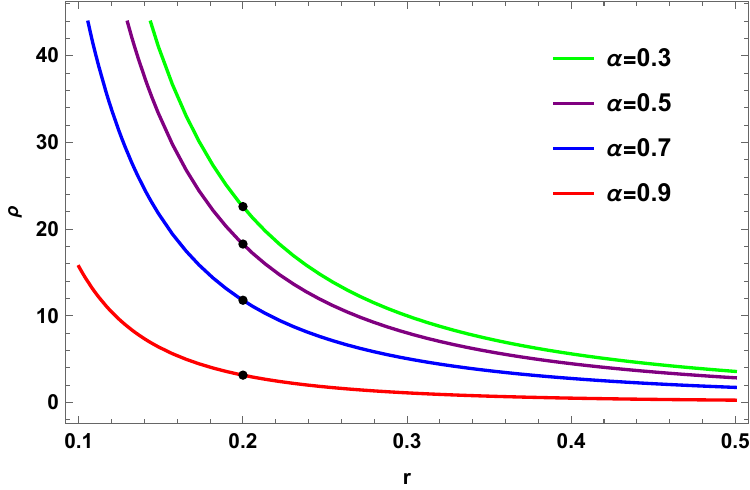}\quad
    \includegraphics[width=0.3\linewidth]{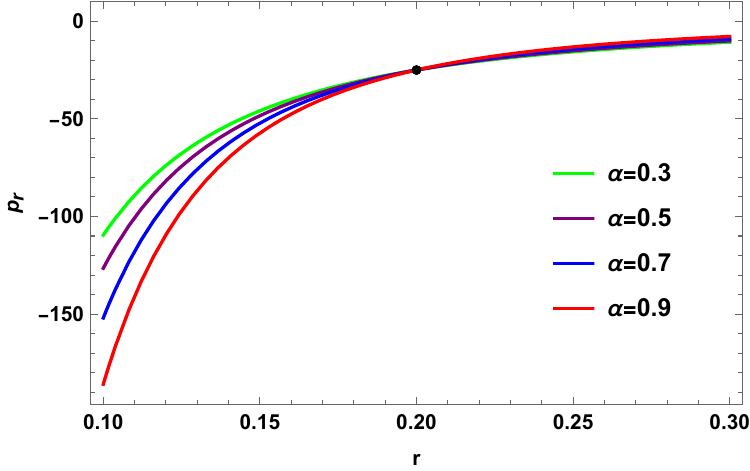}\quad
    \includegraphics[width=0.3\linewidth]{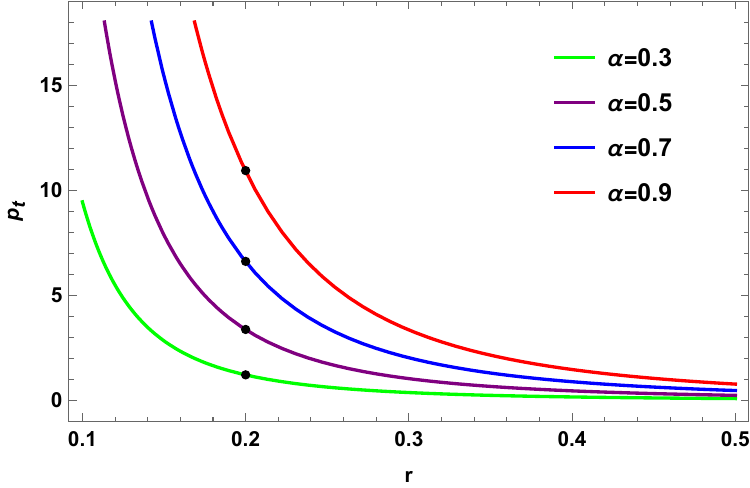}
    \caption{The behaviour of physical quantities $\rho, p_r, p_t$ as a function of $r$ for different values of $\alpha$ for the shape function $A(r)=r_0\,\left(\frac{\cosh r_0}{\cosh r}\right)^{\delta}$ with throat radius $r_0=0.2$ and $\delta=2$.}
    \label{fig:9}
\end{figure*}

and WECs
\begin{eqnarray}
    \text{WEC}_r:&&\rho+p_r=-\frac{\alpha^2\,r_0}{r^2}\,\Big[\delta\,\tanh r+\frac{1}{r} \Big]\,\left(\frac{\cosh r_0}{\cosh r}\right)^{\delta},\nonumber\\
    \text{WEC}_t:&&\rho+p_t=\frac{1-\alpha^2}{r^2}+\frac{\alpha^2\,r_0}{2\,r^2}\,\Big[\frac{1}{r}-\delta\,\tanh r\Big]\,\left(\frac{\cosh r_0}{\cosh r}\right)^{\delta}.\label{mm3}
\end{eqnarray}

The DECs states that
\begin{eqnarray}
    \text{DEC}_r:\rho-|p_r|&=&\frac{1-\alpha^2}{r^2}-\frac{\alpha^2\,\delta\,r_0}{r^2}\,\left(\frac{\cosh r_0}{\cosh r}\right)^{\delta}\,\tanh r\nonumber\\
    &&-\frac{\alpha^2}{r^2}\,\left|1-\frac{1}{\alpha^2}-\frac{r_0}{r}\,\left(\frac{\cosh r_0}{\cosh r}\right)^{\delta}\right|,\nonumber\\
    \text{DEC}_t:\rho-|p_t|&=&\frac{\alpha^2}{r^2}\,\Bigg[\frac{1}{\alpha^2}-1-\delta\,r_0\,\left(\frac{\cosh r_0}{\cosh r}\right)^{\delta}\,\tanh r\Bigg]\nonumber\\
    &&-\frac{\alpha^2\,r_0}{2\,r^2}\,\left|\left(\frac{1}{r}+\delta\,\tanh r\right)\,\left(\frac{\cosh r_0}{\cosh r}\right)^{\delta}\right|.\label{mm4}
\end{eqnarray}

\begin{figure*}[ht!]
    \includegraphics[width=0.3\linewidth]{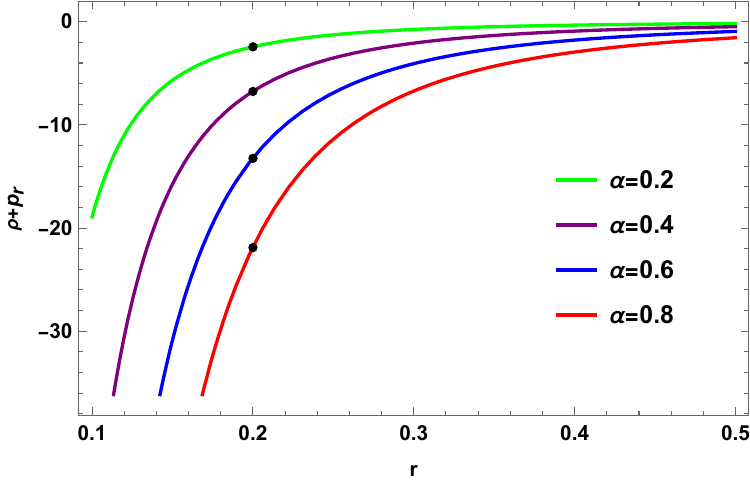}\quad\quad
    \includegraphics[width=0.3\linewidth]{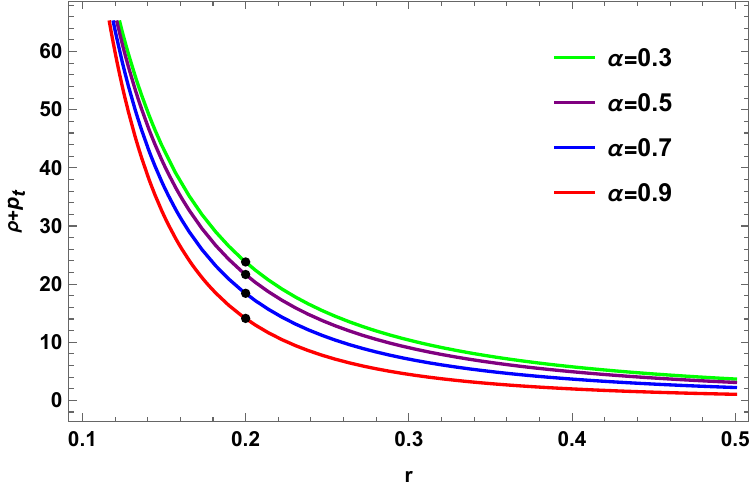}
    \caption{The weak energy condition $\rho+p_r$ (left one) and $\rho+p_t$ (right one) for the shape function $A(r)=r_0\,\left(\frac{\cosh r_0}{\cosh r}\right)^{\delta}$ with throat radius $r_0=0.2$ and $\delta=2$.}
    \label{fig:10}
    \hfill\\
    \includegraphics[width=0.3\linewidth]{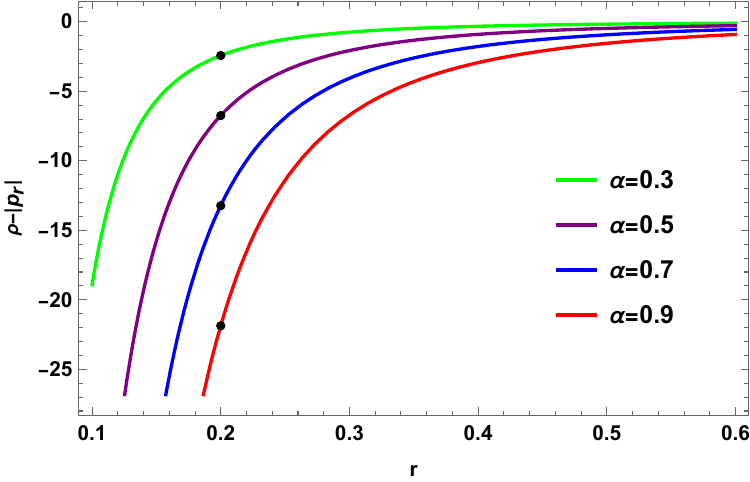}\quad\quad
    \includegraphics[width=0.3\linewidth]{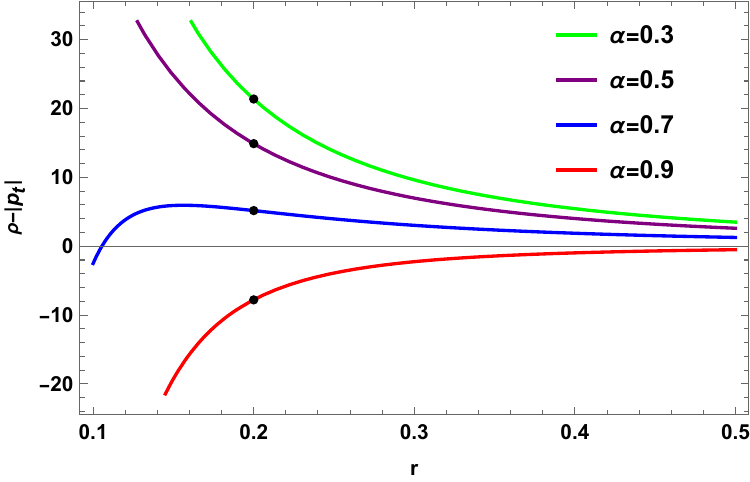}
    \caption{The behaviour of $\rho-|p_r|$ and $\rho-|p_t|$ for the shape function $A(r)=r_0\,\left(\frac{\cosh r_0}{\cosh r}\right)^{\delta}$ with throat radius $r_0=0.2$ and $\delta=2$.}
    \label{fig:11}
\end{figure*}

\begin{figure}[ht!]
    \begin{centering}
    \includegraphics[width=0.6\linewidth]{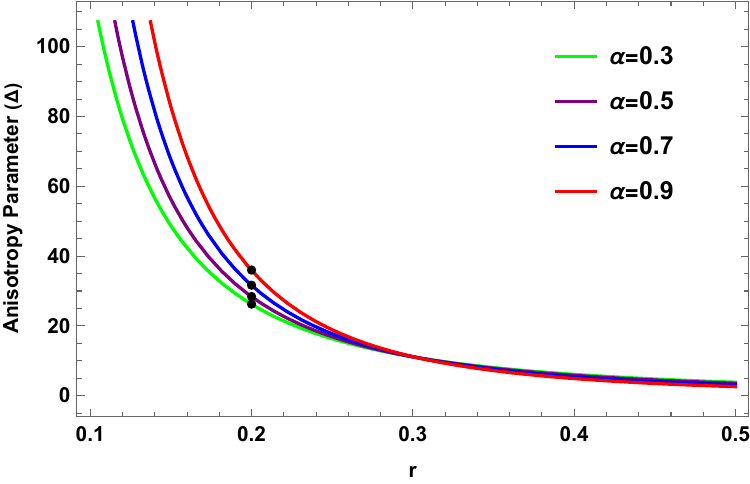}
    \par\end{centering}
    \caption{The nature of aniotropy parameter $\Delta$ as a function of $r$ for the shape function $A(r)=r_0\,\left(\frac{\cosh r_0}{\cosh r}\right)^{\delta}$ with throat radius $r_0=0.2$ and $\delta=2$.}
    \label{fig:12}
\end{figure}

The anisotropy parameter is given by
\begin{equation}
    \Delta=\frac{1-\alpha^2}{r^2}+\frac{\alpha^2\,r_0}{2\,r^2}\,\Bigg(\frac{3}{r}+\delta\,\tanh r \Bigg)\,\left(\frac{\cosh r_0}{\cosh r}\right)^{\delta}.\label{mm5}
\end{equation}

\subsection{Shape Function:\,$A(r)=\frac{1}{r}+\mbox{ln}(r/r_0)$.}\label{subsec:4}

In this model, we consider the following form of the shape function \cite{Chaudhary2024} to examine the topologically charged wormhole space-time given by
\begin{equation}
    A(r)=\frac{1}{r}+\mbox{ln}(r/r_0),\label{nn1}
\end{equation}
where $r_0=1$ is the throat radius. One can show that $A(r=r_0)=r_0=1$.

Therefore, using this shape function (\ref{nn1}), we finds the energy-density, the radial pressure, and the tangential pressure from Eq. (\ref{b7}) as follows:
\begin{eqnarray}
    &&\rho=\frac{1-\alpha^2}{r^2}+\frac{\alpha^2}{r^2}\,\left(-\frac{1}{r^2}+\frac{1}{r}\right),\nonumber\\
    &&p_r=\frac{\alpha^2-1}{r^2}-\frac{\alpha^2}{2\,r^3}\,\left(1/r+\mbox{ln}(r/r_0)\right),\nonumber\\
    &&p_t=\frac{\alpha^2}{2\,r^3}\,\left[2/r+\mbox{ln}(r/r_0)-1\right].\label{nn2}
\end{eqnarray}

\begin{figure*}[ht!]
    \centering
    \includegraphics[width=0.3\linewidth]{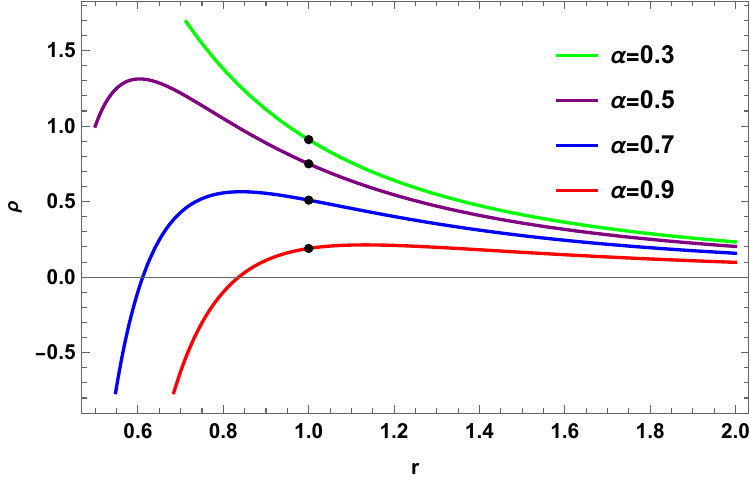}\quad
    \includegraphics[width=0.3\linewidth]{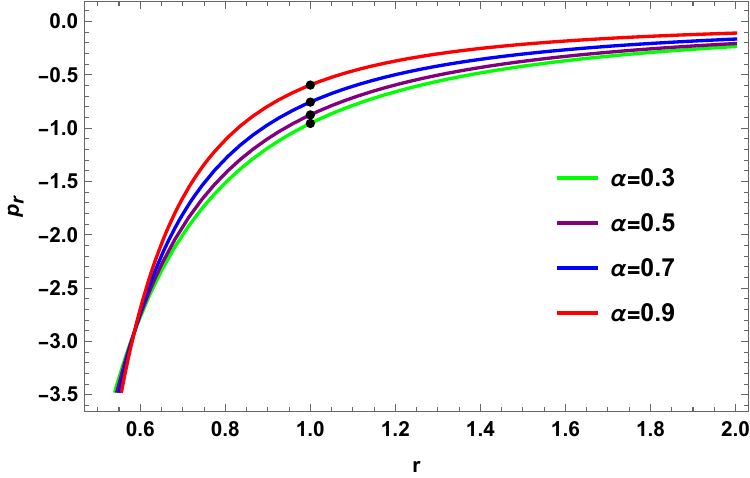}\quad
    \includegraphics[width=0.3\linewidth]{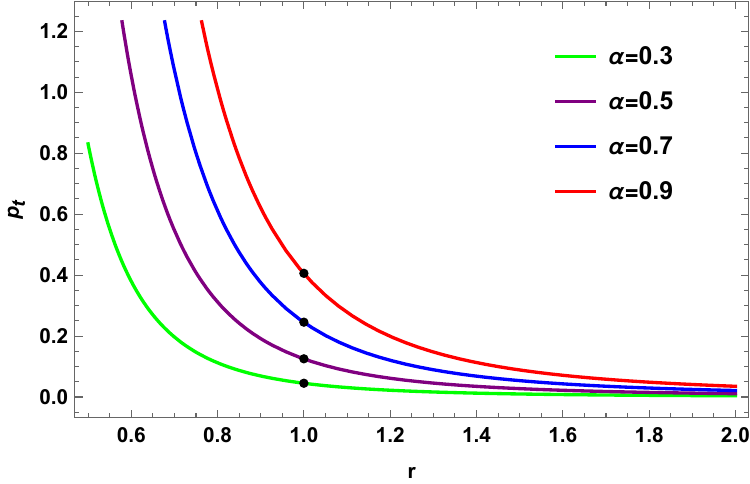}
    \caption{The behaviour of physical quantities $\rho, p_r, p_t$ as a function of $r$ for different values of $\alpha$ for the shape function $A(r)=\frac{1}{r}+\mbox{ln}(r/r_0)$ with throat radius $r_0=0.2$ and $a=0.9<1$.}
    \label{fig:13}
\end{figure*}

\begin{figure*}[ht!]
    \includegraphics[width=0.3\linewidth]{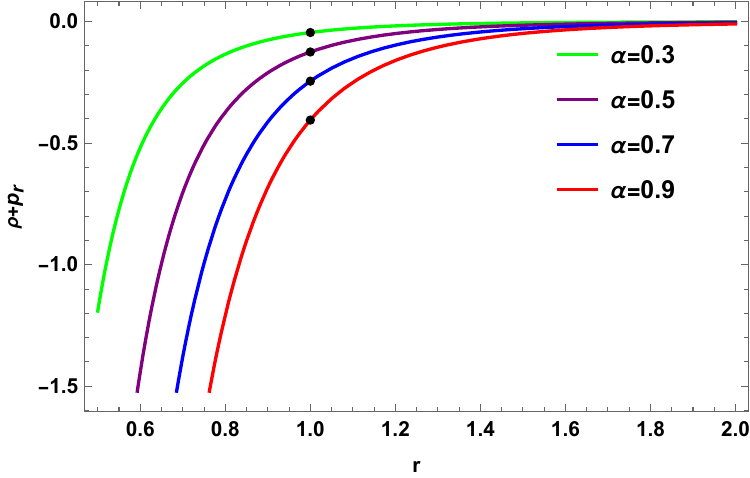}\quad\quad
    \includegraphics[width=0.3\linewidth]{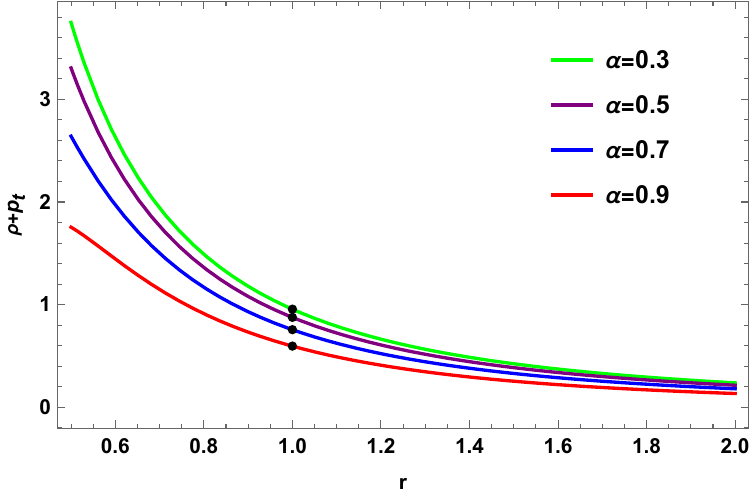}
    \caption{The weak energy condition $\rho+p_r$ (left one) and $\rho+p_t$ (right one) for the shape function $A(r)=\frac{1}{r}+\mbox{ln}(r/r_0)$ with throat radius $r_0=0.2$ and $a=0.9<1$.}
    \label{fig:14}
    \hfill\\
    \includegraphics[width=0.3\linewidth]{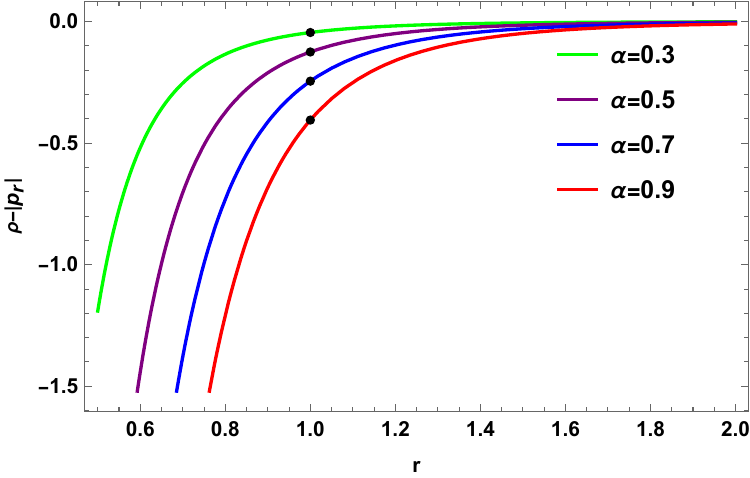}\quad\quad
    \includegraphics[width=0.3\linewidth]{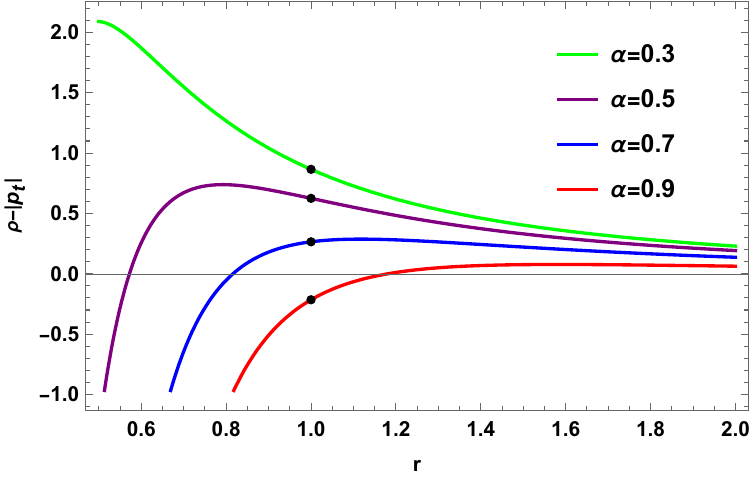}
    \caption{The behaviour of $\rho-|p_r|$ and $\rho-|p_t|$ for the shape function $A(r)=\frac{1}{r}+\mbox{ln}(r/r_0)$ with throat radius $r_0=0.2$ and $a=0.9<1$.}
    \label{fig:15}
\end{figure*}

\begin{figure}[ht!]
    \begin{centering}
    \includegraphics[width=0.6\linewidth]{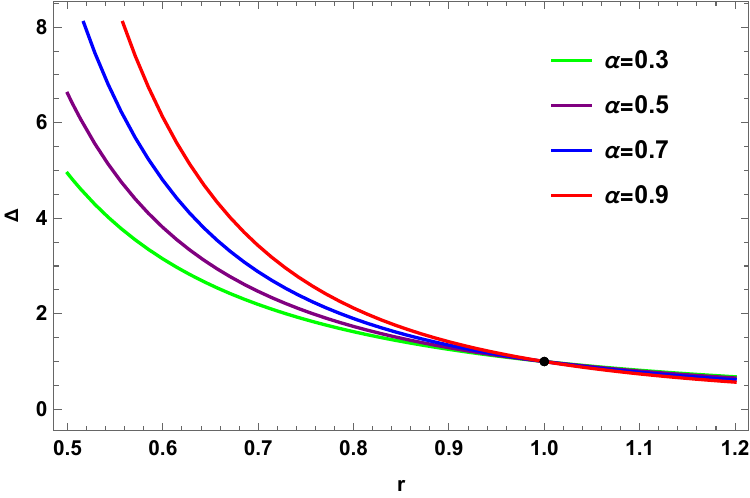}
    \par\end{centering}
    \caption{The nature of anisotropy parameter $\Delta$ as a function of $r$ for the shape function $A(r)=\frac{1}{r}+\mbox{ln}(r/r_0)$ with throat radius $r_0=1.0$ and $a=0.9<1$.}
    \label{fig:16}
\end{figure}

The WECs states that
\begin{eqnarray}
    \text{WEC}_r:\quad &&\rho+p_r=\frac{\alpha^2}{2\,r^3}\,\Bigg[2-\frac{3}{r}-\mbox{ln} (r/r_0) \Bigg],\nonumber\\
    \text{WEC}_t:\quad &&\rho+p_t=\frac{1-\alpha^2}{r^2}+\frac{\alpha^2}{2\,r^3}\,\left[1+\mbox{ln}(r/r_0)\right].\label{nn3}
\end{eqnarray}

The DECs states that
\begin{eqnarray}
    \text{DEC}_r:&&\rho-|p_r|=\frac{1-\alpha^2}{r^2}+\frac{\alpha^2}{r^2}\,\left(-\frac{1}{r^2}+\frac{1}{r}\right)\nonumber\\
    &&-\left|\frac{\alpha^2-1}{r^2}-\frac{\alpha^2}{2\,r^3}\,\left(1/r+\mbox{ln}(r/r_0)\right)\right|,\nonumber\\
    \text{DEC}_t:&&\rho-|p_t|=\frac{1-\alpha^2}{r^2}+\frac{\alpha^2}{r^2}\,\left(-\frac{1}{r^2}+\frac{1}{r}\right)\nonumber\\
    &&-\left|\frac{\alpha^2}{2\,r^3}\,\left[2/r+\mbox{ln}(r/r_0)-1\right]\right|.\label{nn4}
\end{eqnarray}

The anisotropy parameter is given by
\begin{equation}
    \Delta=\frac{1-\alpha^2}{r^2}+\frac{\alpha^2}{2\,r^3}\,\Big[\frac{3}{r}+2\,\mbox{ln} (r/r_0)-1 \Big].\label{nn5}
\end{equation}

\subsection{Shape Function:\,$A(r)=B\,r^n+(1-B)$.}\label{subsec:5}

In this model, we consider the following form of the shape function \cite{Parsaei2020} to examine the topologically charged wormhole space-time given by
\begin{equation}
    A(r)=B\,r^n+(1-B),\label{ss1}
\end{equation}
where $r_0=1.0$ is the throat radius. One can show that the shape function satisfies the relation $A(r=r_0)=r_0=1.0$ for $B$ for any values of $B$.

\begin{figure*}[ht!]
    \centering
    \includegraphics[width=0.3\linewidth]{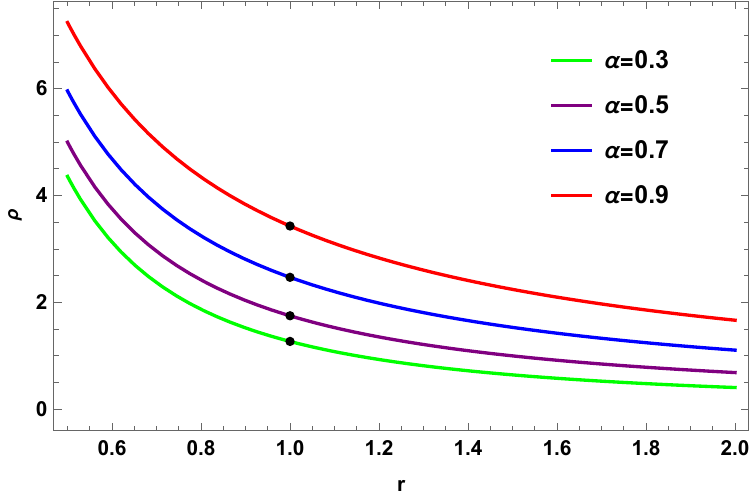}\quad
    \includegraphics[width=0.3\linewidth]{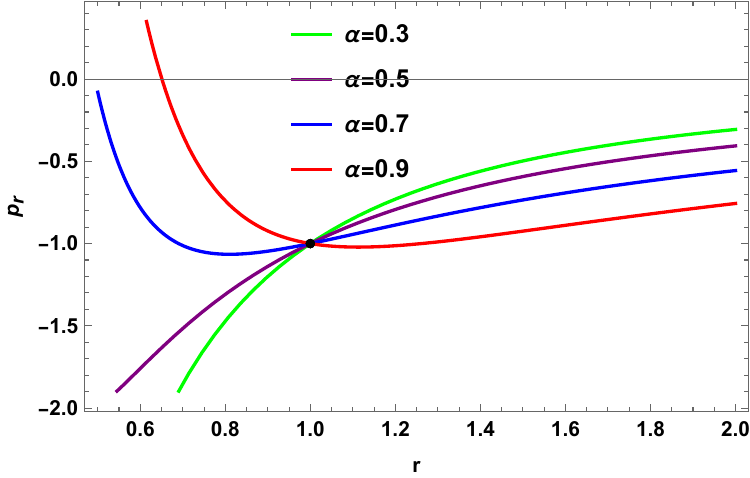}\quad
    \includegraphics[width=0.3\linewidth]{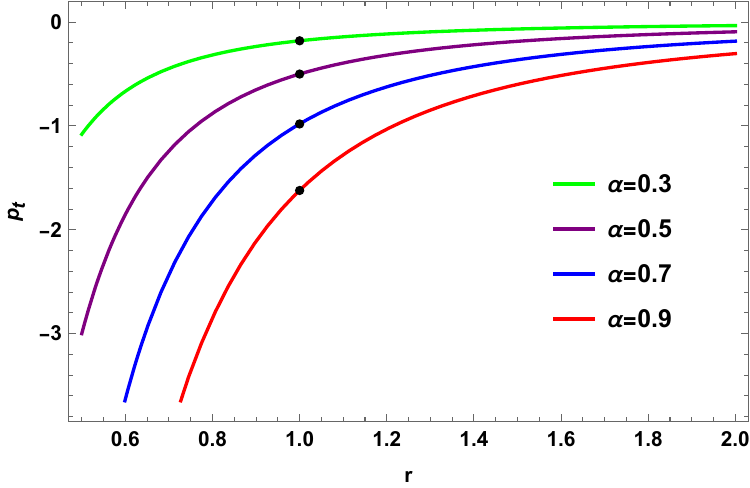}
    \caption{The behavior of physical quantities $\rho, p_r, p_t$ as a function of $r$ for different values of $\alpha$ for the shape function $A(r)=B\,r^n+(1-B)$ with throat radius $r_0=1.0$ and $a=0.9<1$.}
    \label{fig:17}
\end{figure*}

\begin{figure*}[ht!]
    \includegraphics[width=0.3\linewidth]{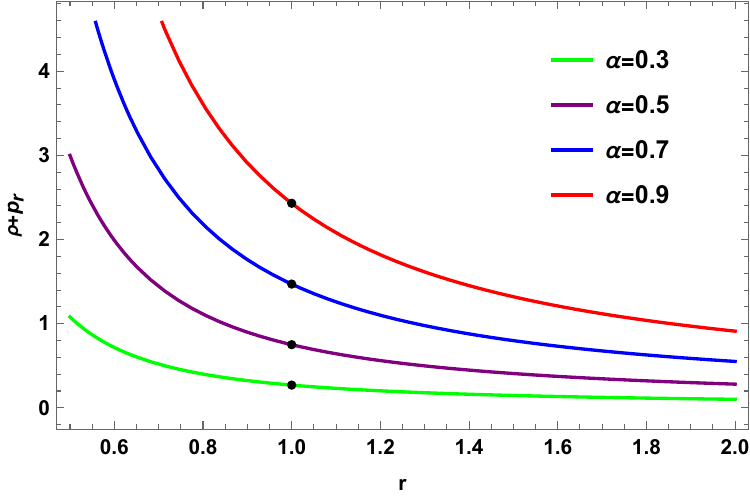}\quad\quad
    \includegraphics[width=0.3\linewidth]{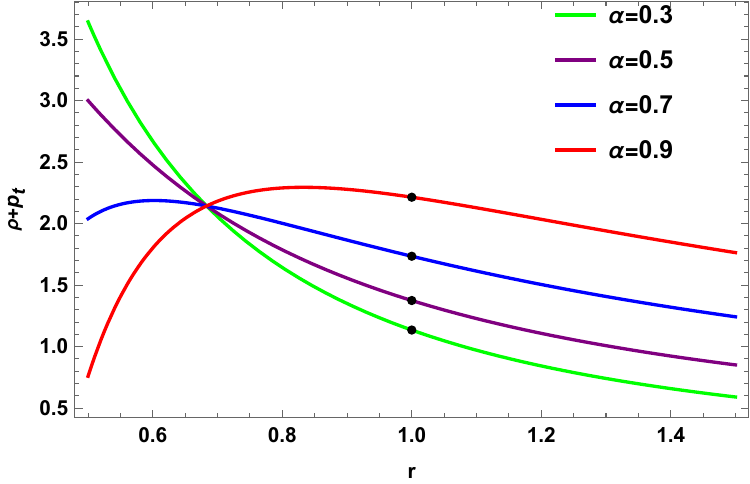}
    \caption{The weak energy condition $\rho+p_r$ (left one) and $\rho+p_t$ (right one) for the shape function $A(r)=B\,r^n+(1-B)$ with throat radius $r_0=0.2$ and $a=0.9<1$.}
    \label{fig:18}
    \hfill\\
    \includegraphics[width=0.3\linewidth]{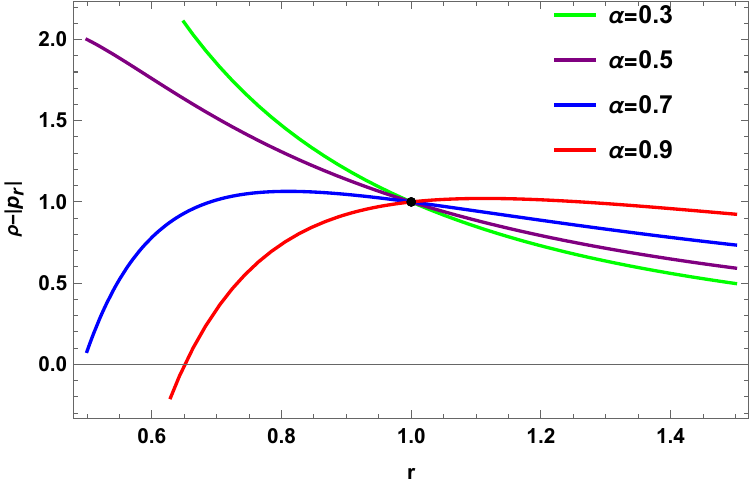}\quad\quad
    \includegraphics[width=0.3\linewidth]{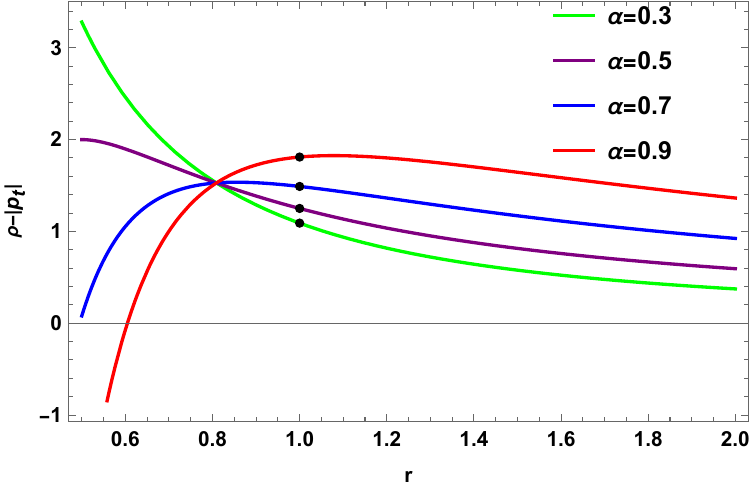}
    \caption{The behavior of $\rho-|p_r|$ and $\rho-|p_t|$ for the shape function $A(r)=B\,r^n+(1-B)$ with throat radius $r_0=0.2$ and $a=0.9<1$.}
    \label{fig:19}
\end{figure*}

\begin{figure}[ht!]
    \begin{centering}
    \includegraphics[width=0.6\linewidth]{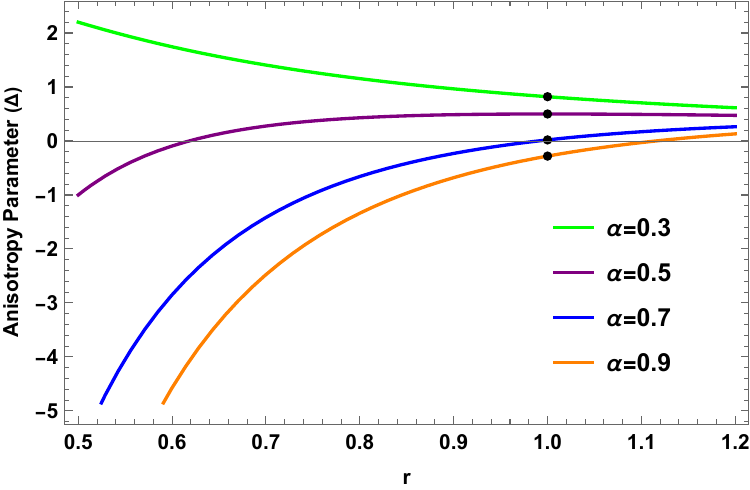}
    \par\end{centering}
    \caption{The nature of anisotropy parameter $\Delta$ as a function of $r$ for the shape function $A(r)=B\,r^n+(1-B)$ with throat radius $r_0=0.2$ and $a=0.9<1$.}
    \label{fig:20}
\end{figure}

Therefore, using this shape function (\ref{ss1}), we finds the energy-density, the radial pressure, and the tangential pressure from Eq. (\ref{b7}) as follows:
\begin{eqnarray}
    &&\rho=\frac{1-\alpha^2}{r^2}+n\,B\,\alpha^2\,r^{n-3},\nonumber\\
    &&p_r=\frac{\alpha^2-1}{r^2}-\frac{\alpha^2}{r^3}\,\left[B\,r^n+(1-B) \right],\nonumber\\
    &&p_t=\frac{\alpha^2}{2\,r^3}\,\left[(1-n)\,B\,r^{n-1}+1-B \right].\label{ss2}
\end{eqnarray}

The WEC states that
\begin{eqnarray}
    \text{WEC}_r:&&\rho+p_r=\alpha^2\,\Big[(n-1)\,B\,r^{n-3}-(1-B)\,r^{-3} \Big],\nonumber\\
    \text{WEC}_t:&&\rho+p_t=\frac{1-\alpha^2}{r^2}+\alpha^2\,\Big[n\,B\,r^{n-3}+\frac{1}{2}\,(1-n)\,B\,r^{n-4}\nonumber\\
    &&+\frac{1}{2}\,(1-B)\,r^{-3}\Big].\label{ss3}
\end{eqnarray}

The DECs states that
\begin{align}
    &\text{DEC}_r:\rho-|p_r|=\frac{1-\alpha^2}{r^2}+n\,B\,\alpha^2\,r^{n-3}\nonumber\\
    &-\left|\frac{\alpha^2-1}{r^2}-\frac{\alpha^2}{r^3}\,\left[B\,r^n+(1-B) \right]\right|,\nonumber\\
    &\text{DEC}_t:\rho-|p_t|=\frac{1-\alpha^2}{r^2}+n\,B\,\alpha^2\,r^{n-3}\nonumber\\
    &-\left|\frac{\alpha^2}{2\,r^3}\,\left[(1-n)\,B\,r^{n-1}+1-B \right]\right|.\label{ss4}
\end{align}

The anisotropy parameter is given by
\begin{equation} 
    \Delta=\frac{1-\alpha^2}{r^2}+\frac{3\,\alpha^2}{2\,r^3}\,(1-B)+\alpha^2\,B\,\Bigg[1+\frac{(1-n)}{2\,r}\Bigg]\,r^{n-3}.\label{ss5}
\end{equation}

\subsection{Shape Function:\,$A(r)=r_0\,\frac{\mbox{ln} (1+r)}{\mbox{ln} (1+r_0)}$.}\label{subsec:6}

In this model, we consider the following form of the shape function \cite{Godani2019} to examine the topologically charged wormhole space-time given by
\begin{equation}
    A(r)=r_0\,\frac{\mbox{ln}(1+r)}{\mbox{ln}(1+r_0)},\label{qq1}
\end{equation}
where $r_0$ is the throat radius.

\begin{figure*}[ht!]
    \centering
    \includegraphics[width=0.3\linewidth]{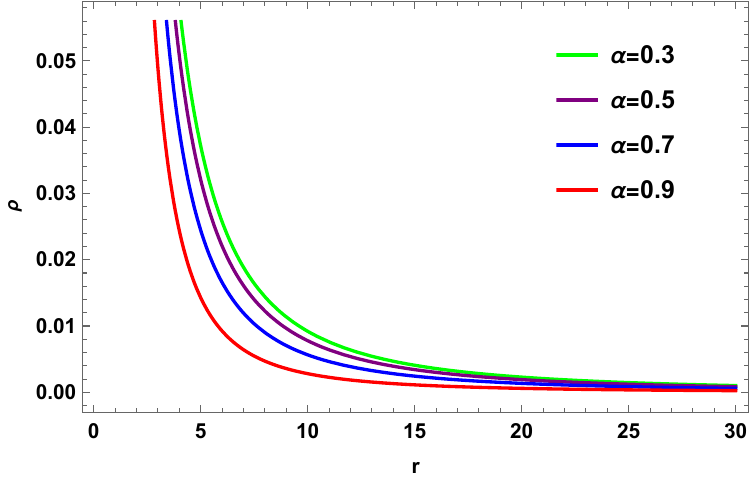}\quad
    \includegraphics[width=0.3\linewidth]{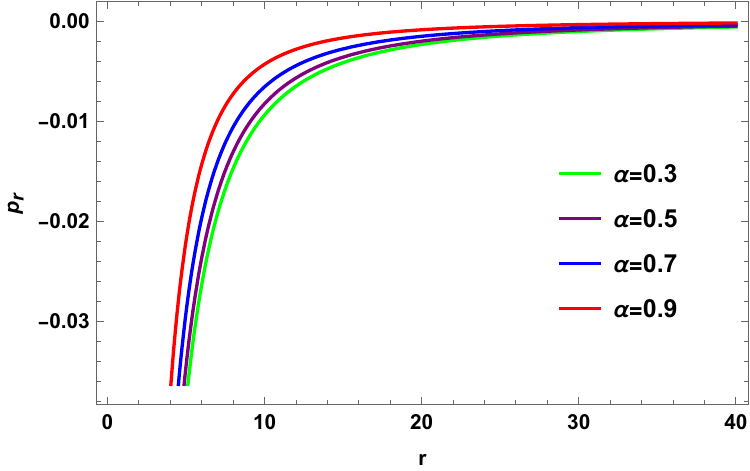}\quad
    \includegraphics[width=0.3\linewidth]{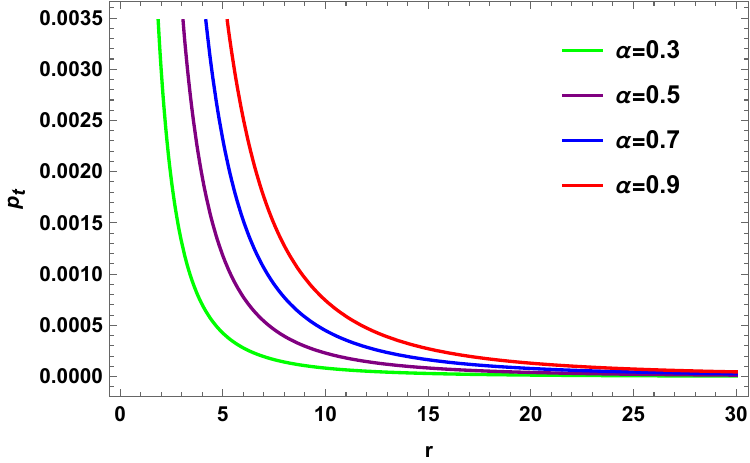}
    \caption{The behavior of physical quantities $\rho, p_r, p_t$ as a function of $r$ for different values of $\alpha$ for the shape function $A(r)=r_0\,\frac{\mbox{ln} (1+r)}{\mbox{ln} (1+r_0)}$ with throat radius $r_0=0.2$ and $a=0.9<1$.}
    \label{fig:21}
\end{figure*}

\begin{figure*}[ht!]
    \includegraphics[width=0.3\linewidth]{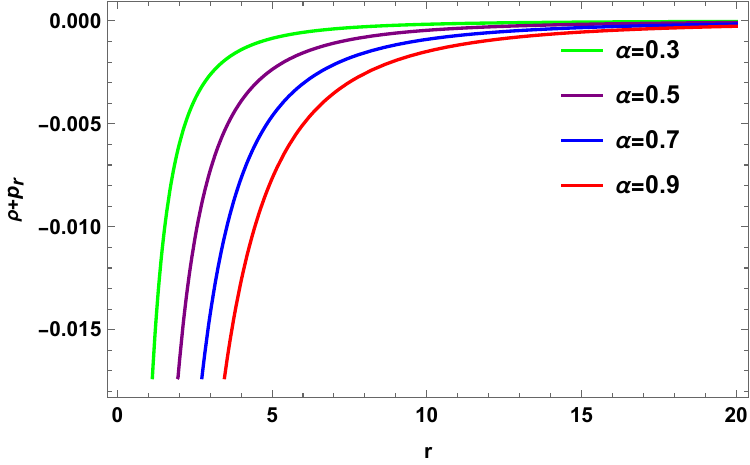}\quad\quad
    \includegraphics[width=0.3\linewidth]{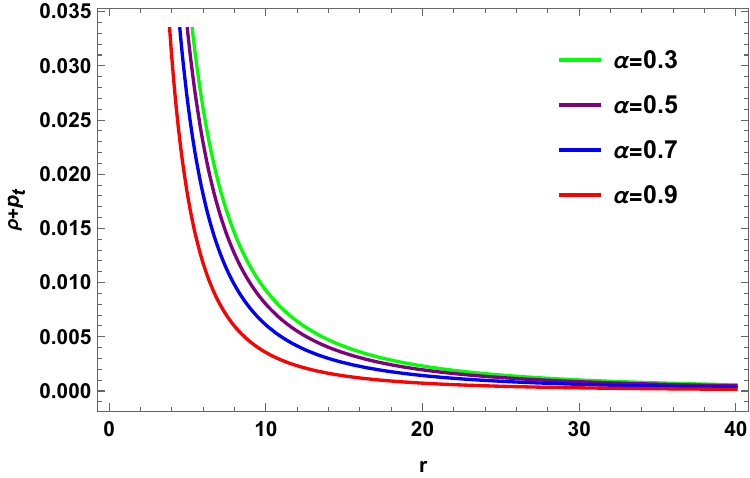}
    \caption{The weak energy condition $\rho+p_r$ (left one) and $\rho+p_t$ (right one) for the shape function $A(r)=r_0\,\frac{\mbox{ln} (1+r)}{\mbox{ln} (1+r_0)}$ with throat radius $r_0=0.2$ and $a=0.9<1$.}
    \label{fig:22}
    \hfill\\
    \includegraphics[width=0.3\linewidth]{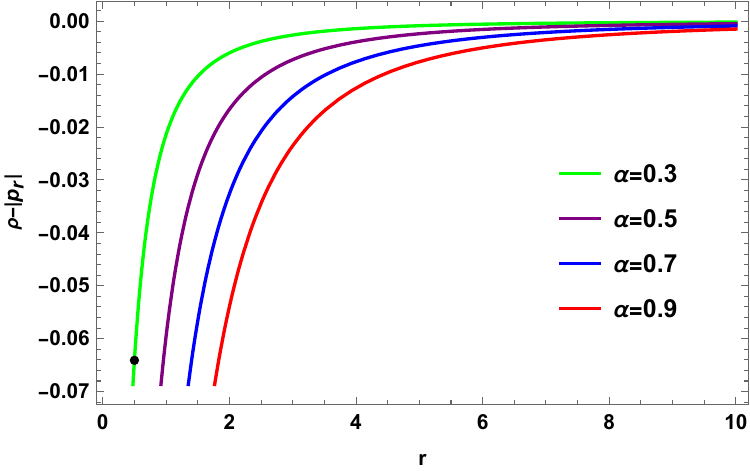}\quad\quad
    \includegraphics[width=0.3\linewidth]{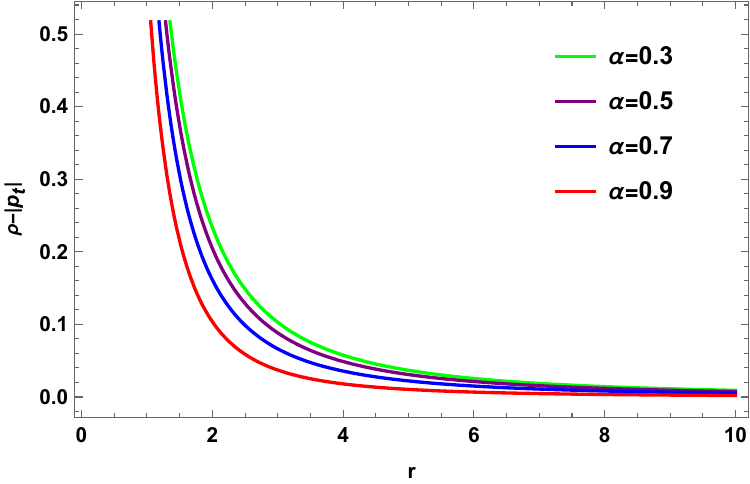}
    \caption{The behavior of $\rho-|p_r|$ and $\rho-|p_t|$ for the shape function $A(r)=r_0\,\frac{\mbox{ln} (1+r)}{\mbox{ln} (1+r_0)}$ with throat radius $r_0=0.2$ and $a=0.9<1$.}
    \label{fig:23}
\end{figure*}

\begin{figure}[ht!]
    \begin{centering}
    \includegraphics[width=0.6\linewidth]{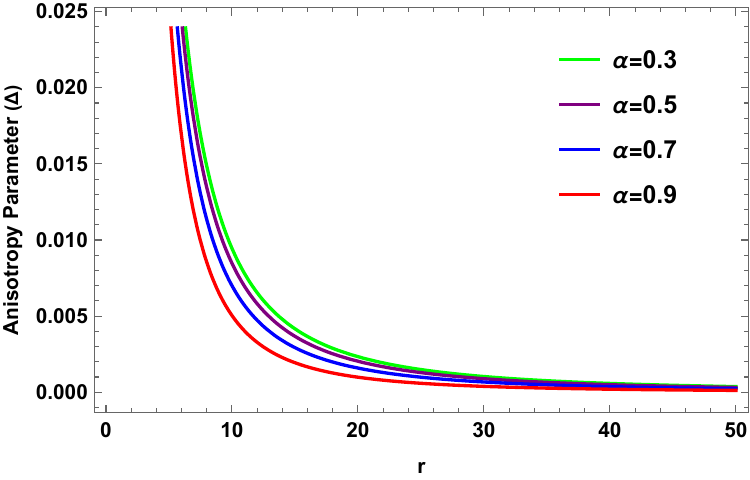}
    \par\end{centering}
    \caption{The nature of anisotropy parameter $\Delta$ as a function of $r$ for the shape function $A(r)=r_0\,\frac{\mbox{ln} (1+r)}{\mbox{ln} (1+r_0)}$ with throat radius $r_0=0.2$ and $a=0.9<1$.}
    \label{fig:24}
\end{figure}

Therefore, using this shape function (\ref{qq1}), we finds the energy-density, the radial pressure, and the tangential pressure from Eq. (\ref{b7}) as follows:
\begin{eqnarray}
    &&\rho=\frac{1-\alpha^2}{r^2}+\frac{\alpha^2\,r_0}{\mbox{ln}(r_0+1)}\,\frac{1}{r^3+r^2},\nonumber\\
    &&p_r=\frac{\alpha^2-1}{r^2}-\frac{\alpha^2\,r_0}{\mbox{ln}(r_0+1)}\,\frac{\mbox{ln}(r+1)}{r^3},\nonumber\\
    &&p_t=\frac{\alpha^2}{2\,r^3}\,\frac{r_0}{\mbox{ln}(r_0+1)}\,\Bigg[\mbox{ln}(r+1)-\frac{r}{r+1}\Bigg].\label{qq2}
\end{eqnarray}

The WECs states that
\begin{eqnarray}
    \text{WEC}_r:\quad &&\rho+p_r=\frac{\alpha^2\,r_0}{\mbox{ln}(r_0+1)}\,\Bigg[\frac{1}{r^2+r^3}-\frac{\mbox{ln}(r+1)}{r^3} \Bigg],\nonumber\\
    \text{WEC}_t:\quad &&\rho+p_t=\frac{1-\alpha^2}{r^2}+\frac{\alpha^2\,r_0}{2\,\mbox{ln}(r_0+1)}\,\Bigg[\frac{1}{r^2+r^3}+\frac{\mbox{ln}(r+1)}{r^3} \Bigg].\quad \label{qq3}
\end{eqnarray}
The DECs states that
\begin{widetext}
\begin{eqnarray}
    \text{DEC}_r:&&\rho-|p_r|=\frac{1-\alpha^2}{r^2}+\frac{\alpha^2\,r_0}{\mbox{ln}(r_0+1)}\,\frac{1}{r^3+r^2}-\left|\frac{\alpha^2-1}{r^2}-\frac{\alpha^2\,r_0}{\mbox{ln}(r_0+1)}\,\frac{\mbox{ln}(r+1)}{r^3}\right|,\nonumber\\
    \text{DEC}_t:&&\rho-|p_t|=\frac{1-\alpha^2}{r^2}+\frac{\alpha^2\,r_0}{\mbox{ln}(r_0+1)}\,\frac{1}{r^3+r^2}-\left|\frac{\alpha^2}{2\,r^3}\,\frac{r_0}{\mbox{ln}(r_0+1)}\,\Bigg[\mbox{ln}(r+1)-\frac{r}{r+1}\Bigg]\right|.\label{qq4}
\end{eqnarray}
\end{widetext}

The anisotropy parameter is given by
\begin{equation}
    \Delta=\frac{1-\alpha^2}{r^2}+\frac{3\,\alpha^2\,r_0}{2\,r^2}\,\frac{1}{\mbox{ln}(r_0+1)}\,\Big[\frac{\mbox{ln}(r+1)}{r}-\frac{1}{r+1}  \Big].\label{qq5}
\end{equation}

\subsection{Shape Function:\,$A(r)=r_0+a\,r_0\,\Big[\Big(\frac{r}{r_0}\Big)^{\beta}-1\Big]$.}\label{subsec:7}

In this model, we consider the following form of the shape function \cite{Lobo2013} to examine the topologically charged wormhole space-time given by
\begin{equation}
    A(r)=r_0+a\,r_0\,\Big[\Big(\frac{r}{r_0}\Big)^{\beta}-1\Big],\label{kk1}
\end{equation}
where $r_0$ is the throat radius and we have the following restrictions on the parameters:
\begin{equation}
    \beta<1,\quad\quad\quad 0 < a\,\beta <1.\label{kk2}
\end{equation}

\begin{figure*}[ht!]
    \centering
    \includegraphics[width=0.3\linewidth]{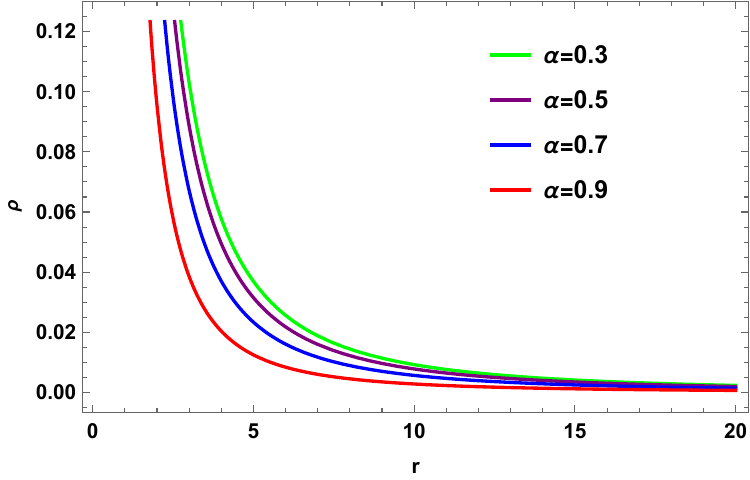}\quad
    \includegraphics[width=0.3\linewidth]{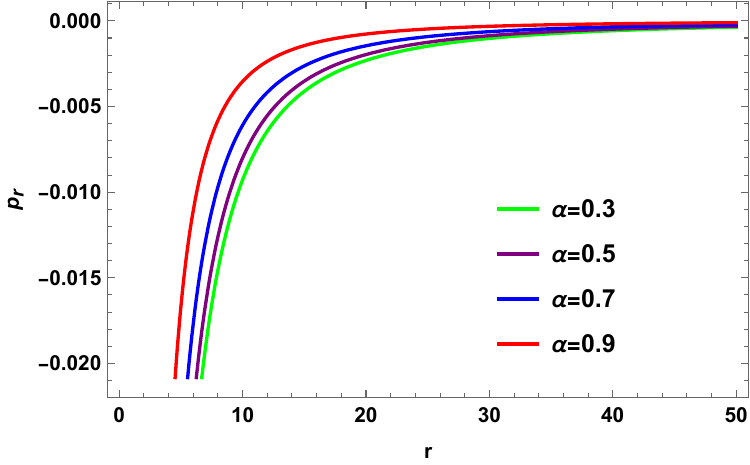}\quad
    \includegraphics[width=0.3\linewidth]{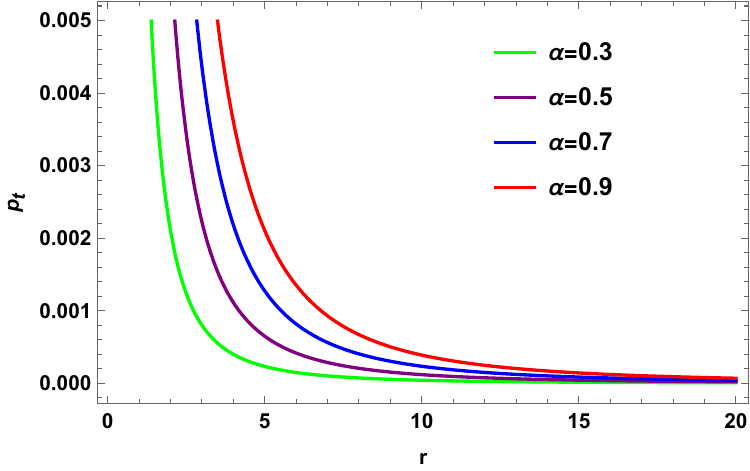}
    \caption{The behavior of physical quantities $\rho, p_r, p_t$ as a function of $r$ for different values of $\alpha$ for the shape function $A(r)=r_0+a\,r_0\,\Big[\Big(\frac{r}{r_0}\Big)^{\beta}-1\Big]$ with throat radius $r_0=0.2$ and $a=0.9<1$.}
    \label{fig:25}
\end{figure*}

\begin{figure*}[ht!]
    \includegraphics[width=0.3\linewidth]{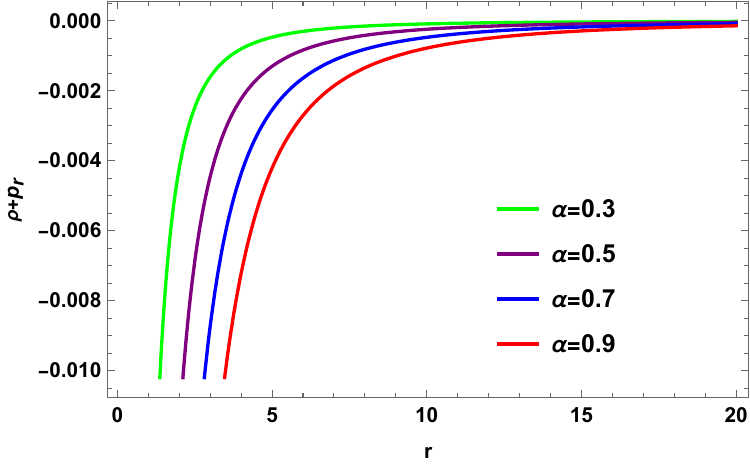}\quad\quad
    \includegraphics[width=0.3\linewidth]{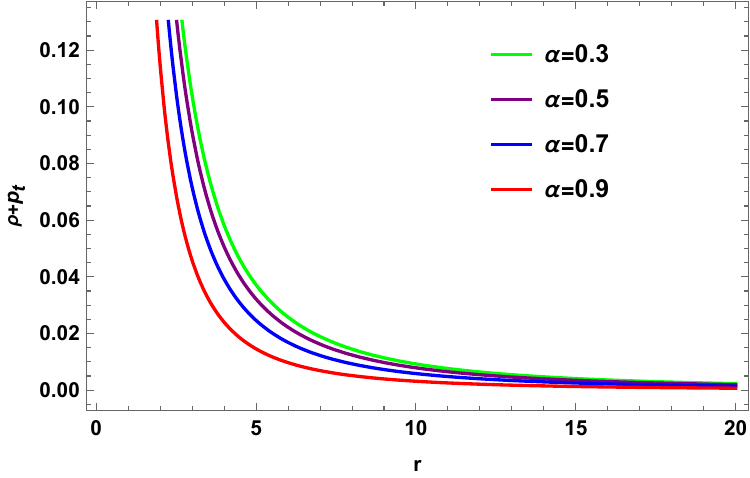}
    \caption{The weak energy condition $\rho+p_r$ (left one) and $\rho+p_t$ (right one) for the shape function $A(r)=r_0+a\,r_0\,\Big[\Big(\frac{r}{r_0}\Big)^{\beta}-1\Big]$ with throat radius $r_0=0.2$ and $a=0.9<1$.}
    \label{fig:26}
    \hfill\\
    \includegraphics[width=0.3\linewidth]{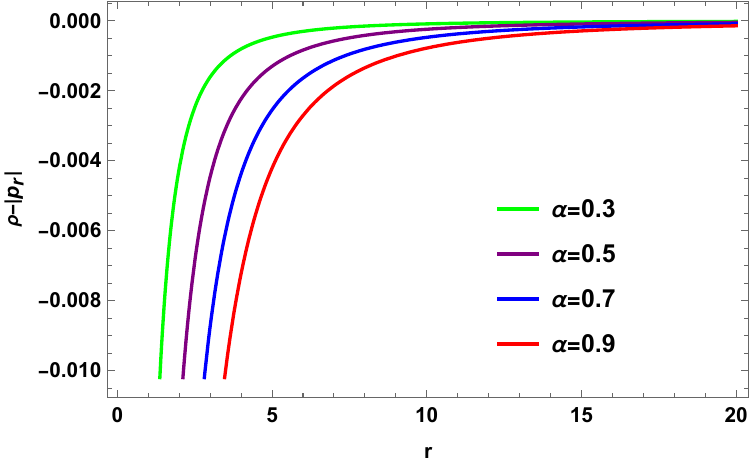}\quad\quad
    \includegraphics[width=0.3\linewidth]{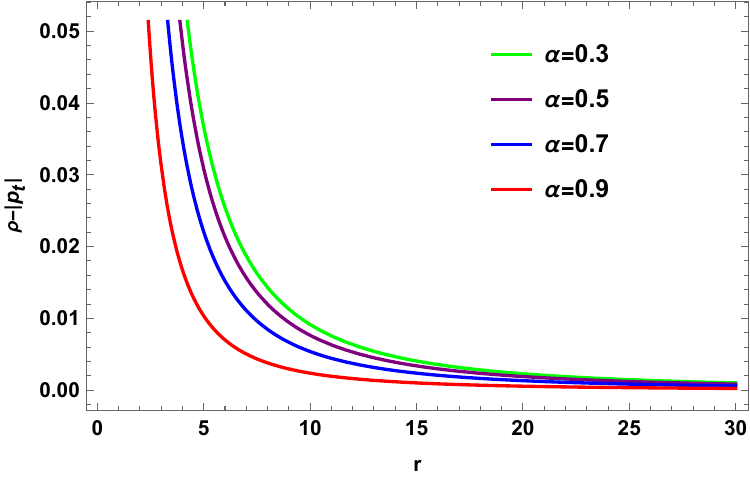}
    \caption{The behavior of $\rho-|p_r|$ and $\rho-|p_t|$ for the shape function $A(r)=r_0+a\,r_0\,\Big[\Big(\frac{r}{r_0}\Big)^{\beta}-1\Big]$ with throat radius $r_0=0.2$ and $a=0.9<1$.}
    \label{fig:27}
\end{figure*}

\begin{figure}[ht!]
    \begin{centering}
    \includegraphics[width=0.6\linewidth]{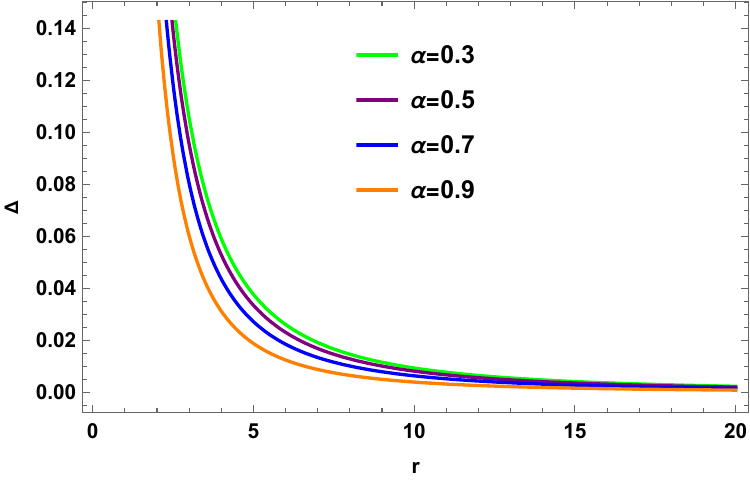}
    \par\end{centering}
    \caption{The nature of anisotropy parameter $\Delta$ as a function of $r$ for the shape function $A(r)=r_0+a\,r_0\,\Big[\Big(\frac{r}{r_0}\Big)^{\beta}-1\Big]$ with throat radius $r_0=0.2$ and $a=0.9<1$.}
    \label{fig:28}
\end{figure}

Therefore, using this shape function (\ref{kk1}), we finds the energy-density, the radial pressure, and the tangential pressure from Eq. (\ref{b7}) as follows:
\begin{eqnarray}
    &&\rho=\frac{1-\alpha^2}{r^2}+\frac{\alpha^2\,a\,\beta}{r^2}\,\Big(\frac{r}{r_0}\Big)^{\beta-1},\nonumber\\
    &&p_r=\frac{\alpha^2-1}{r^2}-\frac{\alpha^2}{r^3}\,\Bigg[r_0+a\,r_0\,\Big\{\Big(\frac{r}{r_0}\Big)^{\beta}-1\Big\}\Bigg],\nonumber\\
    &&p_t=\frac{\alpha^2}{2\,r^3}\,\Bigg[(1-a)\,r_0+a\,r_0\,(1-\beta)\,\Big(\frac{r}{r_0}\Big)^{\beta} \Bigg].\label{kk3}
\end{eqnarray}

The WEC states that
\begin{widetext}
\begin{eqnarray}
    \text{WEC}_r:\quad &&\rho+p_r=\frac{\alpha^2}{r^2}\,(a\,\beta-a)\,\Big(\frac{r}{r_0}\Big)^{\beta-1}+\alpha^2\,(a-1)\,\frac{r_0}{r^3},\nonumber\\
    \text{WEC}_t:\quad &&\rho+p_t=\frac{1-\alpha^2}{r^2}+\alpha^2\,(1-a)\frac{r_0}{2\,r^3}+\frac{\alpha^2\,(a\,\beta+a)}{2\,r^2}\,\Big(\frac{r}{r_0}\Big)^{\beta-1}.\label{kk4}
\end{eqnarray}
\end{widetext}
The DECs states that
\begin{eqnarray}
\text{DEC}_r:&&\rho-|p_r|=\frac{1-\alpha^2}{r^2}+\frac{\alpha^2\,a\,\beta}{r^2}\,\Big(\frac{r}{r_0}\Big)^{\beta-1}\nonumber\\
&&-\left|\frac{\alpha^2-1}{r^2}-\frac{\alpha^2}{r^3}\,\Bigg[r_0+a\,r_0\,\Big\{\Big(\frac{r}{r_0}\Big)^{\beta}-1\Big\}\Bigg]\right|,\nonumber\\
\text{DEC}_t:&&\rho-|p_t|=\frac{1-\alpha^2}{r^2}+\frac{\alpha^2\,a\,\beta}{r^2}\,\Big(\frac{r}{r_0}\Big)^{\beta-1}\nonumber\\
&&-\left|\frac{\alpha^2}{2\,r^3}\,\Bigg[(1-a)\,r_0+a\,r_0\,(1-\beta)\,\Big(\frac{r}{r_0}\Big)^{\beta} \Bigg]\right|.\label{kk5}
\end{eqnarray}

The anisotropy parameter is given by
\begin{equation}
    \Delta=\frac{1-\alpha^2}{r^2}+\frac{\alpha^2\,r_0}{2\,r^3}\,\Bigg[3\,(1-a)+(3\,a-a\,\beta)\,\Big(\frac{r}{r_0}\Big)^{\beta}\Bigg].\label{kk6}
\end{equation}

\section{Results and Discussion}

In this section, we summarizes our study of the wormhole space-time featuring a global monopole by selecting various shape functions.

\begin{enumerate}

    \item In Figure \ref{fig:1}, we illustrate the behavior of energy-density $\rho$, the radial $p_r$ and tangential pressure $p_t$ of the fluid as a function of $r$ for different values of the global monopole parameter $\alpha$. Here , we have chosen the wormhole throat radius $r_0=0.2$. We have seen that the energy-density of the fluid is positive. 
    
In Figures \ref{fig:2} to \ref{fig:3}, the weak energy condition and the dominant energy condition were generated as a function of $r$ for $\alpha=0.6,0.7,0.8,0.9$ and the throat radius $r_0=0.2$. We have observed that both these energy conditions violate along the radial direction and satisfied along the tangential direction.

We have generated this ansiotropy parameter in Figure \ref{fig:4} as a function of $r$ for different values of the global monopole parameter $\alpha$. 

Thus, we can conclude that matter-energy content in the wormhole model with shape function $A(r)=r_0\,\exp(r_0-r)$ discussed in \ref{subsec:1} featuring a global monopole have positive energy density. However, this model partially violated the weak and dominant energy conditions along the radial direction and satisfied along the tangential direction.

\item We have plotted the relevant physical quantities in Figure~\ref{fig:5} to illustrate their dependence on the radial coordinate $r$ for different values of the global monopole parameter $\alpha$. It is observed that the energy density ($\rho$) and tangential pressure ($p_t$) remain positive throughout the spacetime, whereas the radial pressure ($p_r$) is negative.

Figures~\ref{fig:6} and \ref{fig:7} depict the variation of the quantities $\rho \pm p_r$ and $\rho \pm p_t$ with respect to $r$ for different values of $\alpha$. The plots show that these combinations remain positive over the entire domain considered. Since the energy density is also positive, the matter distribution satisfies both the weak energy condition (WEC) and the dominant energy condition (DEC).

Figure~\ref{fig:8} illustrates the behaviour of the anisotropy parameter as a function of $r$. It is evident that the anisotropy parameter is positive in the vicinity of the wormhole throat ($r=r_0=0.2$), indicating that the anisotropic force is directed outward. Consequently, the wormhole geometry exhibits a repulsive nature, which helps to counterbalance the gravitational attraction and contributes to the stability of the wormhole structure.

Therefore, we conclude that the wormhole model characterized by the shape function $A(r)=r_0\,\frac{a^r}{a^r_{0}}$ in \ref{subsec:2} with a global monopole charge, is supported by an anisotropic matter distribution with positive energy density. Furthermore, the model satisfies both the weak and dominant energy conditions throughout the spacetime. These results indicate that the proposed global monopole wormhole configuration represents a physically viable and well-behaved wormhole solution.

\item In Figure~\ref{fig:13}, we have depicted the energy density, the radial pressure and the tangential pressure as a function of $r$ for different values of the global monopole parameter $\alpha$. We see that the energy density and tangential pressure remain positive near the throat while the radial pressure becomes negative.

In Figures~\ref{fig:14} to \ref{fig:15}, we have plotted the weak and dominant energy conditions along the radial and tangential directions by varying $\alpha$. From these Figures, it is observed that the conditions $\rho \pm p_r$ is violated while the condition $\rho \pm p_t$ satisfied at the wormhole throat $r=r_0=1.0$.

In Figure~\ref{fig:16}, we have plotted the anisotropy parameter is positive near the throat $r=r_0=1.0$. This indicates the wormhole metric with global monopole charge is repulsive in nature.

Thus, we can conclude that matter-energy content anisotropic fluid in the wormhole model with shape function $A(r)=1/r+\ln\!(r/r_0)$ featuring a global monopole in \ref{subsec:3} charge have positive energy density. Moreover, this model partially satisfied both the weak and the dominant energy conditions at the throat and is repulsive in nature.

\item  In Figure~\ref{fig:9}, we have depicted the relevant physical quantities as a function of $r$ for different values of the global monopole parameter $\alpha=0.3, 0.5, 0.7, 0.9$. It is observed that the energy density ($\rho$) and tangential pressure ($p_t$) remain positive near the throat, whereas the radial pressure ($p_r$) is negative.

In Figure~\ref{fig:10}, we have illustrated the weak energy condition as a function of $r$ for different values of the global monopole parameter $\alpha$. From the Figure, it is evident that the weak energy condition $(\rho+p_r)$ violates, whereas $(\rho-p_t)$ satisfies at the wormhole throat $r=r_0=0.2$.

In Figure~\ref{fig:11}, we have plotted the strong energy condition as a function of $r$ by varying $\alpha$. We see that this energy condition is partially satisfied along the tangential direction provided small values of the global monopole parameter $\alpha \leq 0.7$.

In Figure~\ref{fig:12}, we have shown that the anisotropy parameter is positive near the throat $r=r_0=0.2$. This indicates the wormhole metric with global monopole charge is repulsive in nature.
 
Thus, we can conclude that matter-energy content anisotropic fluid in the wormhole model with the shape function $A(r)=r_0\,\left(\frac{\cosh r_0}{\cosh r}\right)^{\delta}$ where $\delta=2$ featuring a global monopole charge in \ref{subsec:4} have positive energy density. Moreover, this model partially satisfied both the weak and the dominant energy conditions and is repulsive in nature.

\item In Figure~\ref{fig:17}, we have depicted the energy density, the radial pressure and the tangential pressure as a function of $r$ for different values of the global monopole parameter $\alpha$. We see that the energy density remain positive, while the tangential and radial pressures near the throat are negative.

In Figures~\ref{fig:18} to \ref{fig:19}, we have depicted the weak and dominant energy conditions along the radial and tangential directions by varying $\alpha$. From these Figures, it is evident that the condition $\rho \pm p_r$ and $\rho \pm p_t$ are satisfied at the wormhole throat $r=r_0=1.0$.

In Figure~\ref{fig:20}, we have shown that the anisotropy parameter is positive near the throat $r=r_0=1.0$ provided for small values of global monopole parameter $\alpha$. This indicates the wormhole space-time with global monopole under the considered shape function charge is repulsive in nature for lower values of $\alpha$ and attractive for higher values of $\alpha$.

Thus, we can conclude that matter-energy content anisotropic fluid in the wormhole model with shape function $A(r)=B r^n+1-B$ featuring a global monopole charge in \ref{subsec:5} have positive energy density. Moreover, this model partially satisfied both the weak and the dominant energy conditions at the throat and is repulsive in nature for smaller values of global monopole parameter.

\item In Figure~\ref{fig:21}, we have depicted the energy density, the radial pressure and the tangential pressure as a function of $r$ for different values of the global monopole parameter $\alpha$. We see that the energy density and tangential pressure are positive near the throat, while the radial pressure is negative.

In Figures~\ref{fig:22} to \ref{fig:23}, we have plotted the weak and dominant energy conditions along the radial and tangential directions by varying $\alpha$. From these Figures, it is evident that the condition $\rho \pm p_r$ is negative, while the condition $\rho \pm p_t$ is positive at the wormhole throat $r=r_0=0.2$.

In Figure~\ref{fig:24}, we have shown that the anisotropy parameter is positive near the throat $r=r_0=0.2$. This indicates the wormhole metric with global monopole charge is repulsive in nature.

Thus, we can conclude that matter-energy content anisotropic fluid in the wormhole model with shape function $A(r)=r_0\,\frac{\mbox{ln} (1+r)}{\mbox{ln} (1+r_0)}$ featuring a global monopole charge in \ref{subsec:6} have positive energy density. Moreover, this model partially satisfied both the weak and the dominant energy conditions at the throat and is repulsive in nature.

\item In Figure~\ref{fig:25}, we have depicted the energy density, the radial pressure and the tangential pressure as a function of $r$ for different values of the global monopole parameter $\alpha$. We see that the energy density and tangential pressure remain positive near the throat while the radial pressure becomes negative.

In Figures~\ref{fig:26} to \ref{fig:27}, we have generated the weak and dominant energy conditions along the radial and tangential directions for the aforementioned global monopole parameter. From the Figures, it is evident that the condition $\rho \pm p_r$ is violated while the condition $\rho \pm p_t$ satisfied at the wormhole throat $r=r_0=0.2$.

In Figure~\ref{fig:28}, we have shown that the anisotropy parameter is positive near the throat $r=r_0=0.2$. This indicates the wormhole metric with global monopole charge is repulsive in nature.

Thus, we can conclude that matter-energy content anisotropic fluid in the wormhole model with shape function $A(r)=r_0+a\,r_0\,\Big[\Big(\frac{r}{r_0}\Big)^{\beta}-1\Big]$ featuring a global monopole charge in \ref{subsec:7} have positive energy density. Moreover, this model partially satisfied both the weak and the dominant energy conditions at the throat and is repulsive in nature.

\end{enumerate}

\section{Conclusions}\label{conclusions}

The comprehensive study of a class of Morris-Thorne-type wormholes by incorporating a global monopole parameter and examining energy conditions associated with various shape functions enhances our understanding of their potential existence in the universe. While the speculation about the necessity of exotic matter with negative energy to sustain wormholes remains a key focus, the pursuit of such elusive elements and the fascination with the prospect of space-time shortcuts continues to captivate both physicists and science fiction enthusiasts. This work, therefore, contributes significantly to the ongoing discourse on the theoretical possibilities and implications of these enigmatic passages in the fabric of space-time, especially in the presence of global monopole charge. Considering a constant redshift function, we obtain exact expressions for the energy density, radial and tangential pressures, and the anisotropy parameter for seven physically motivated shape functions, and investigate energy conditions, including NEC, WEC, SEC, and DEC. Our results demonstrate that the inclusion of a global monopole charge parameter substantially alters the physical properties of traversable wormholes. For all the shape functions considered, the energy density remains positive throughout spacetime, providing a more physically acceptable matter distribution than in the conventional Morris-Thorne traversable wormhole. The energy conditions are strongly fulfilled by the choice of the shape functions. While a few of the shape functions partially violate the energy conditions, namely, NEC, WEC, and DEC, especially along the radial directions, other functions completely satisfy the energy conditions throughout the spacetime, thereby diminishing the need for the exotic matter field.

Among the various wormhole geometries considered here, the shape functions
\[
A(r)=r_0\,e^{-(r-r_0)},\quad A(r)=r_0\,\frac{a^r}{a^r_{0}},\quad A(r)=B r^n+1-B
\]
emerges as the most physically acceptable solutions. The rationale behind such a choice is reflected through the derivation of the positive energy density, which subsequently leads to the fulfillment of both the NEC and DEC. These demonstrate that the presence of a global monopole can support a traversable wormhole with an ordinary anisotropic matter source. The remaining shape functions comprise intermediate cases in which the violation of energy conditions is sufficiently reduced in comparison to the conventional Morris-Thorne wormholes. Nevertheless, the anisotropy parameter reveals that the topological charge enhances the effective gravitational attraction inside the wormhole. Depending on the form of the shape functions, the anisotropic stresses may lead to the creation of either repulsive or attractive wormhole configurations. These functions greatly influence the spacetime topology in determining the internal structure and stability of wormhole geometries.   

Our results suggest that the topological charge parameter offers a promising avenue for constructing more physically viable traversable wormholes. They improve the matter content by providing a positive energy density, and for suitable shape functions, sufficiently reduce the violations of the energy conditions. These findings provide strong support to consider the spacetime topology as a promising mechanism in shaping wormhole physics and motivate us for further investigations of topologically charged traversable wormholes in modified theories of gravity. In the future, we will examine the stability of these wormholes using linear perturbations, QNMs, gravitational lensing, and shadows to assess their distinguishability from black holes using present-era precision gravitational astronomy.

\section*{Acknowledgments}

F.A. and M. S. A. acknowledge the Inter University Centre for Astronomy and Astrophysics (IUCAA), Pune, India, for granting a visiting associateship.

\section*{Data Availability Statement}

No new data were generated in this study.

\section*{Conflicts of Interest}

Authors declare no conflicts of interest.

\appendix

\section{Energy conditions and formation of wormhole throat}
A wormhole arises from a special solution to Einstein’s field equations
\begin{equation}
    G_{\mu\nu}+\Lambda\,g_{\mu\nu}=T_{\mu\nu},\label{a1}
\end{equation}
where $G_{\mu\nu}=(R_{\mu\nu}-1/2\,R\,g_{\mu\nu})$ is the Einstein tensor, $g_{\mu\nu}$ is the metric tensor, $\Lambda$ is the cosmological constant, $T_{\mu\nu}$ is the energy-momentum tensor, $R_{\mu\nu}$ is the Ricci tensor, and $R=g_{\mu\nu}\,R^{\mu\nu}$ is the Ricci scalar.

Energy conditions are the mathematical constraints that dictate how matter and energy behave within space- time. In general relativity, all physically realizable energy-momentum tensors are expected to fulfill the energy conditions. The different energy conditions are as follows \cite{Hawking1973,Wald1984}:

\begin{itemize}
    
\item The Null energy condition (NEC): $\rho+p_i \geq 0\quad  \forall \quad i \in [1,2,3]$.

\item The Weak energy condition (WEC): $\rho \geq 0$ and $\rho+p_i \geq 0\quad \forall\quad  i \in [1,2,3]$.

\item The Strong energy condition (SEC): $\rho+p_i+2\,p_j \geq 0\quad \forall\quad i,j \in [1,2,3]$.

\item The Dominate energy condition (DEC): $\rho-|p_i| \geq 0\quad \forall\quad i \in [1,2,3]$.
    
\end{itemize}

Anisotropy means that the properties of the system vary with direction. Thus, in general relativity, the anisotropy parameter refers to quantities that measure the degree of anisotropy or directional dependence of a physical system or space-time geometry. It is defined by
\begin{equation}
    \Delta=p_t-p_r,\label{a2}
\end{equation}
where $p_r$ and $p_t$ are the radial and tangential pressures, respectively. The geometry is attractive or repulsive if $\Delta <0$ or $\Delta>0$, respectively. For $\Delta=0$, the fluid content in the wormhole is isotropic.

The line element for a circularly symmetric and static four-dimensional traversable wormhole space-time in the Schwarzschild coordinate is given by \cite{MorrisThorne1988,Morris1988}
\begin{equation}
    ds^2=-e^{2\,\Phi(r)}\,dt^2+\frac{dr^2}{\left(1-\frac{A(r)}{r}\right)}+r^2\,d\theta^2+r^2\,\sin^2\theta\,d\phi^2.\label{a3} 
\end{equation}
The radial coordinate $r$ has a range that increases from a minimum value at $r_0$, corresponding to the radius of the wormhole throat, to $\infty$, i.e., $r \in [r_0,\infty)$ and other coordinates are $-\infty < t < +\infty$, $0 \leq \theta < \pi$, and $0 \leq \phi \leq 2\,\pi$. Here $\Phi(r)$ is the redshift function and $A(r)$ is the shape function. The shape function $A(r)$ in line-element (\ref{a7}) must satisfy flare-out condition \cite{MorrisThorne1988,Morris1988,Visser1995}. At the wormhole throat, the shape function $A(r)$ must satisfy the condition
\begin{equation}
    A(r)|_{r=r_0}=r_0.\label{a4}
\end{equation}
Although the metric is singular at $r=r_0$, proper distance as an invariant quantity must be well-behaved, and therefore, the following integral must be real and regular out the throat \cite{MorrisThorne1988,Morris1988,Visser1995},
\begin{equation}
    \ell(r)=\int^{\infty}_{r_0}\,\frac{dr}{\sqrt{1-A(r)/r}}.\label{a5}
\end{equation}
And for $r>r_0$, i.e., for beyond the wormhole throat, it complies with the condition
\begin{equation}
    A(r)/r <1.\label{a6}
\end{equation}

\begin{itemize}
    \item The shape function $A(r)$ must satisfy the flaring out condition at $r=r_0$, i.e.,
    \begin{equation}
    A'(r=r_0) <1.\label{a7}
    \end{equation}
    Here, superscript $(')$ denotes ordinary radial derivative.

    \item The condition for the asymptotically flat space-time geometry is given by
    \begin{equation}
       \mbox{As}\quad r \to \infty,\quad \frac{A}{r} \to 0\quad \mbox{and}\quad \Phi(r) \to 0.\label{a8} 
    \end{equation}

    \item The redshift function $\Phi(r)$ ought to be finite and must not vanish at the throat $r_0$.
    
\end{itemize}

\bibliographystyle{apsrev4-2}
\bibliography{reference}

\end{document}